\documentclass[Journal]{IEEEtran}
\usepackage{amsmath,amsthm, mathtools}
\usepackage[noadjust]{cite} 

\usepackage{graphicx}
\usepackage{epstopdf}
\usepackage{color}
\usepackage{amsfonts,amsmath,amssymb}
\usepackage{cite}
\usepackage{url}
\usepackage{bm}
\usepackage{bbm}
\usepackage{amssymb}
\usepackage{pdfpages}
\def\BibTeX{{\rm B\kern-.05em{\sc i\kern-.025em b}\kern-.08em T\kern-.1667em\lower.7ex\hbox{E}\kern-.125emX}}
\usepackage{fancyhdr}
\fancypagestyle{distribution}{
	\fancyhead{}
	\fancyfoot{}
	\fancyfoot[L]{\tiny Approved for Public Release; Distribution is Unlimited; $\#$20-1996; Dated 10/13/2020.}
}
\fancypagestyle{nothing}{
	\fancyhead{}
	\fancyfoot{}
}
\usepackage{caption}
\usepackage{subcaption}
\DeclareCaptionFont{mysize}{\fontsize{8}{9.6}\selectfont}
\newcommand{\beq}{\begin{equation}}
\newcommand{\eeq}{\end{equation}}
\newcommand{\ket}[1]{\left\lvert #1 \right>}
\newcommand{\bra}[1]{\left< #1 \right\rvert}
\newcommand{\ketbra}[3]{\left\lvert #1  \right> _{#3}\left< #2\right\rvert}
\newcommand{\braket}[2]{\left< #1 | #2 \right>}

\newcommand*{\bigchi}{\raisebox{2pt}{\mbox{\Large$\chi$}}}

\newcommand{\tr}{\textrm{Tr}}
\newcommand{\ao}{\alpha_0}
\newcommand{\ea}{{\epsilon_{A'}}}
\newcommand{\eb}{{\epsilon_{B'}}}

\begin{document}
	\bstctlcite{IEEEexample:BSTcontrol}

\title{Satellite-based Distribution of Hybrid Entanglement}
\author{Hung Do}
\author{
    \IEEEauthorblockN{Hung Do\IEEEauthorrefmark{1}, Robert Malaney\IEEEauthorrefmark{1} and Jonathan Green\IEEEauthorrefmark{2}}\\
    \IEEEauthorblockA{\IEEEauthorrefmark{1}School of Electrical Engineering and Telecommunications,\\ The University of New South Wales, Sydney, NSW 2052, Australia.}\\
    \IEEEauthorblockA{\IEEEauthorrefmark{2}Northrop Grumman Corporation, San Diego, California, USA.}
}

\maketitle
\thispagestyle{distribution} 
\renewcommand{\headrulewidth}{0pt}

\begin{abstract}

Heterogeneous quantum networks consisting of
mixed-technologies - Continuous Variable (CV) and Discrete
Variable (DV) - will become ubiquitous
as global quantum communication matures. Hybrid
quantum-entanglement between CV and
DV modes will be a critical resource in such networks.
A leading candidate for such  hybrid quantum entanglement is that between
Schr\"odinger-cat states and photon-number states. In this work,
we explore, for the first time, the use of Two-Mode Squeezed Vacuum (TMSV)
states, distributed from satellites, as a teleportation resource for the
re-distribution of our candidate hybrid entanglement
pre-stored within terrestrial quantum networks.
We determine the loss conditions under which  teleportation via the
TMSV resource outperforms direct-satellite
distribution of the hybrid entanglement, in addition to
quantifying the advantage of teleporting the DV mode
relative to the CV mode. 
Our detailed calculations show that under the loss conditions anticipated from Low-Earth-Orbit, DV teleportation via the TMSV resource will always provide for significantly improved outcomes, relative to other means for distributing hybrid entanglement within heterogeneous quantum networks.
\end{abstract}

\section{Introduction}
%


The Schr\"odinger-cat state is of fundamental importance because it represents the superposition of two macroscopic quantum states. In the optical domain, the cat state  can take the form of a superposition of two coherent states of opposite phase \cite{dodonov1974even}.  Such a cat state finds applications in many areas including universal quantum computing via Continuous Variables (CV), where the quantum information is encoded in the quadratures of the optical field \cite{ralph2003quantum,lund2008fault,brask2010ahybrid,sangouard2010quantumrepeaters}. CV protocols can in many instances be  more efficient relative to Discrete Variable (DV) versions of the same protocols, where the quantum information is encoded in the DV properties of single photons such as polarization or photon-number \cite{park2010entangled,morin2013remote,takeda2015entanglementswapping}.

\begin{figure}[h!]
    \centering
		\begin{subfigure}[b]{0.95\linewidth}
        \includegraphics[width=\linewidth]{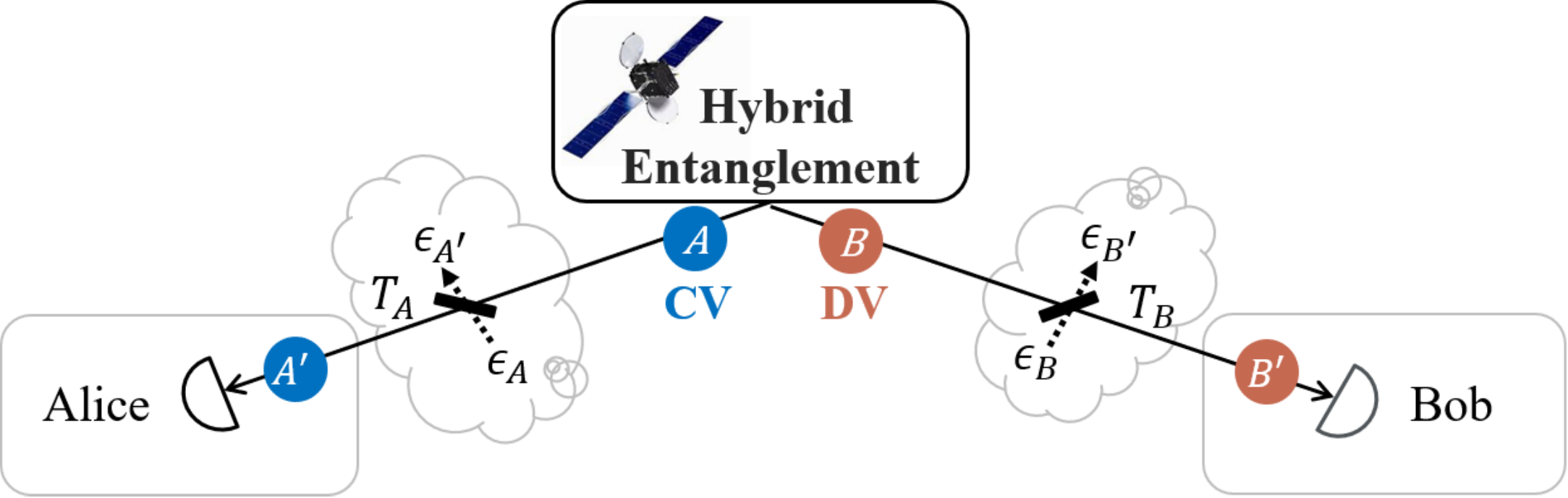}
        \caption{Direct distribution.}
        \label{direct_setup}
				\vspace{2em}
    \end{subfigure}
		\begin{subfigure}[b]{1\linewidth}
        \includegraphics[width=\linewidth]{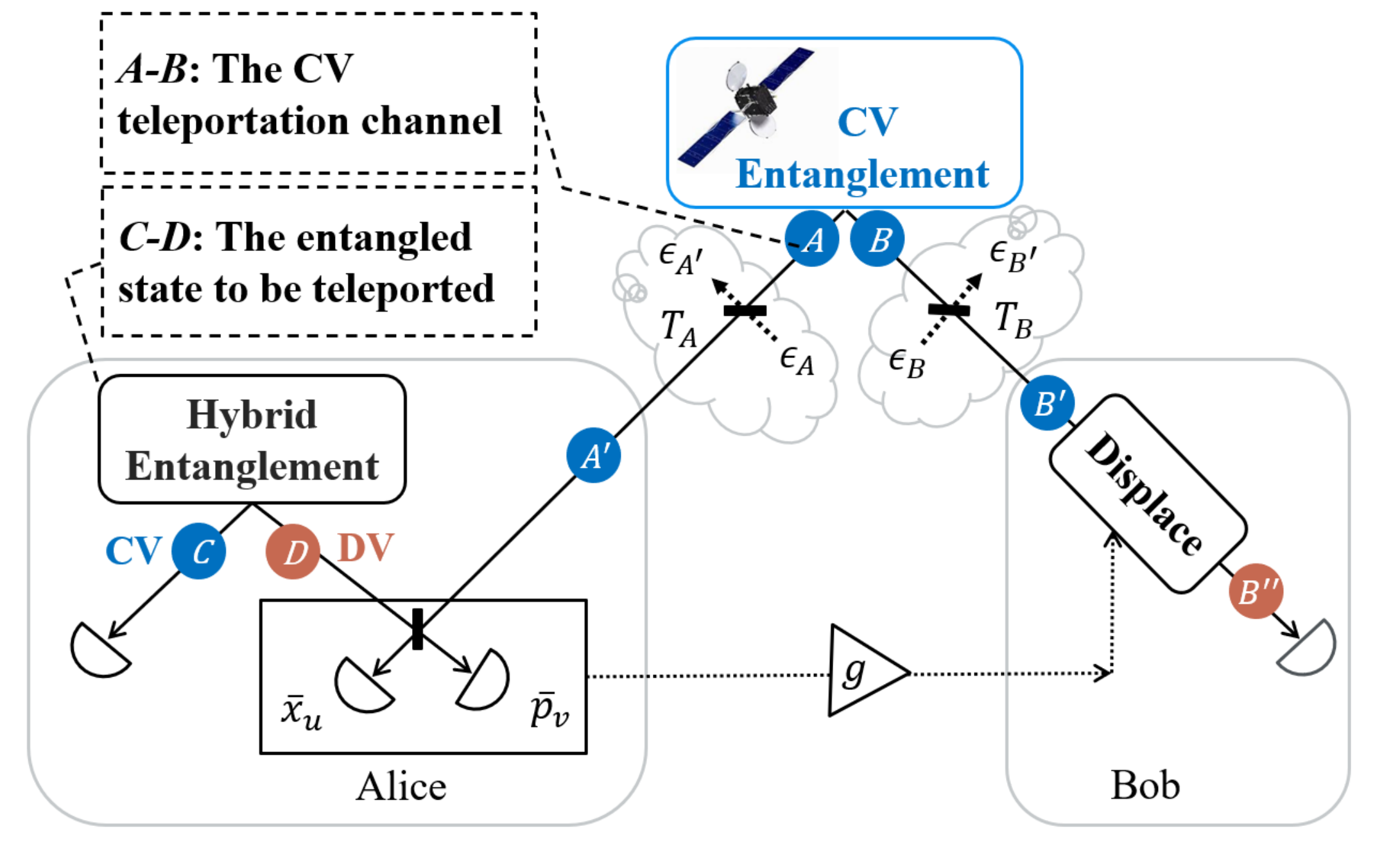}
        \caption{Teleportation.}
        \label{teleported_setup}
    \end{subfigure}
    \caption{(a) The hybrid entangled state $A- B$ is beamed directly from the satellite down to Earth. (b) In the hybrid scheme, the attenuated CV entanglement of the modes $A'-B'$ is used as a teleportation channel. The hybrid entangled state $C-D$ is comprised of the CV mode $C$ and the DV mode $D$. Mode $D$ (or $C$) is teleported through the CV teleportation channel, with teleportation gain $g$, to create the final mode $B''$ entangled to $C$ (or $D$). ${\bar x_u}$ and ${\bar p_v}$ stand for the Bell state measurement outputs. In both plots, $\epsilon_A$ and $\epsilon_B$  represent the vacuum contributions, while $T_A$ and $T_B$ represent the channel transmissivities. }
\label{g_opt_E_max}
\end{figure}

Hybrid entanglement between quantum states can be considered  as the entanglement between two physically separated modes, one of which is encoded in DV information while the other is encoded in CV information. An example of such a hybrid state would be entanglement between the CV Schr\"odinger-cat states and the DV photon-number states.
Such a hybrid entangled state has been demonstrated to have applications
in quantum control, specifically for the remote preparation of a CV qubit using a local DV mode \cite{morin2013remote,huang2013performance,lejeannic2018remote,sychez2018entanglement}. In quantum communications, hybrid entangled states can lead to the violation of the steering inequality  \cite{cavailles2018demonstration}, thus generating a positive key rate for the one-sided device-independent Quantum Key Distribution (QKD) protocol  \cite{branciard2012onesided}. Impressively, the hybrid entangled state can be used to distribute secret keys over 1000km of optical fibre via the coherent-state-based twin-field QKD protocol \cite{yin2019coherent, yin2019finite, zhang2019improving}.

Hybrid entanglement may become particularly important for long-distance communication via satellites, especially as a mechanism to interconnect (via teleportation and entanglement swapping) terrestrial devices operating within heterogeneous (mixed DV and CV) networks. Connecting such mixed-technology devices through traditional fiber links has its limitations since the loss in optical fiber scales exponentially with distance.

In contrast, for a satellite in Low-Earth-Orbit (LEO), which is about 500km above the ground, the Micius experiment has shown that the down-link (satellite to ground) channel is largely affected by atmospheric turbulence and diffraction \cite{yin2017satellite}.
In future systems, this level of loss will likely be further mitigated by large receiving telescopes and/or  adaptive optical tracking techniques \cite{zhao2012aberration,li2014evaluation}. It is, therefore, of great interest to explore the use of entanglement distribution via satellites as a means to interconnect terrestrial devices running on mixed technologies.

Previous work has studied the scenario where both the state to be teleported and the teleportation channel are hybrid entangled states \cite{sychez2018entanglement,lim2016loss,ulanov2017teleportation,guccione2020connecting}. However, the generation of a hybrid-entangled teleportation resource channel is experimentally challenging - especially when the source is placed on board a satellite as in Fig.~\ref{direct_setup}. 
On the other hand, the generation of CV entanglement in the form of a Two-Mode Squeezed Vacuum (TMSV) state is relatively easy to achieve and potentially available as a satellite-based technology \cite{yin2017satellite}. In this work, we study, for the first time, the use of a TMSV channel (as shown in Fig.~\ref{teleported_setup}) to teleport a hybrid entangled state.  

The use of the attenuated TMSV channel for teleportation (henceforth simply referred to as the teleportation) was previously
studied using the Wigner function formalism \cite{braustein1998teleportationCV}, the P-function formalism \cite{takeoka2003continuous}, and the Fock basis formalism \cite{lie2019limitations}. However, such mathematical models were limited to the teleportation of common single-mode states, such as the vacuum state, the single-photon state \cite{lie2019limitations}, the coherent state \cite{takeoka2003continuous}, or the Schr\"odinger's cat state \cite{braustein1998teleportationCV}.
The Wigner function formalism was then extended to describe the teleportation of one mode in a \emph{pure} DV entangled state, which takes into consideration the cross-diagonal matrices of zero and single photon numbers, $\ketbra{0}{1}{}$ and $\ketbra{1}{0}{}$ \cite{takeda2013gaintuning,do2019hybrid}. Different from \cite{braustein1998teleportationCV,takeoka2003continuous, lie2019limitations,takeda2013gaintuning, do2019hybrid}, in this work, we describe the teleportation of either the DV or CV mode of \emph{hybrid} entanglement. Since the DV mode is only comprised of \emph{zero} and \emph{single} photon numbers, the DV-mode teleportation can be derived by extending the Wigner-function formalism  in \cite{takeda2013gaintuning,do2019hybrid}. The CV mode, however, is comprised of cross-diagonal matrices of \emph{multi}-photon numbers, where the teleportation is too complex to be described by the previous formalisms. 
As a solution, we use the characteristic function formalism to
model the CV-mode teleportation, identifying a new simplification in this formalism 
that allows for analytic determination of the teleportation fidelity of a hybrid state.

Hybrid entanglement distribution will be important for the
design of future heterogeneous quantum networks. Our two main novel
contributions to this end, relative to the existing literature, are:
\begin{itemize}
\item We determine the teleportation fidelity of a hybrid state
when either the DV or CV mode is passed through a
lossy TMSV channel.

\item We calculate the fidelity of a hybrid entangled state
following a direct distribution of each mode through a
differential lossy channel; using this calculation to then
quantify the difference in performance between the direct 
and teleported distribution schemes as a function of loss in the 
satellite-to-ground channel.

\end{itemize}

The structure of the remainder of this paper is as follows. Section~\ref{hybrid_entanglement_math} introduces hybrid entanglement. Section~\ref{direct_distribution} studies the direct distribution of hybrid entanglement, with transmission loss in both the DV and CV mode. Section~\ref{teleportation} studies the teleportation of either mode of the hybrid entanglement via a lossy TMSV channel. Section~\ref{result} compares the results between CV/DV-mode teleportation and direct distribution, highlighting the different advantages of both schemes.
Section~\ref{conclusion} summarizes our findings.
\section{Hybrid entanglement}
\label{hybrid_entanglement_math}
Consider the hybrid entanglement between cat states and qubit states
\begin{align}
&\ket{\psi}_{h} = \frac{\ket{cat_-}_C\ket{0}_D + \ket{cat_+}_C\ket{1}_D}{\sqrt{2}},
\label{hybrid_ket}
\end{align}
where for clarity, we have specified the spatial modes as $C$ and $D$ following the setup in Fig.~\ref{teleported_setup}. $\ket{0}$ and $\ket{1}$ are the vacuum state and the single photon state in the photon-number basis, while the Schr\"{o}dinger cat states are given by
\beq
\ket{cat_\pm} = \frac{\ket{\alpha_0} \pm \ket{-\alpha_0}}{N_{\pm}},
\label{cat}
\eeq
with $\ket{\pm \alpha_0}$ denoting the coherent states of 
amplitude $|\alpha_0|$, and the normalization constants being
\beq
N_{\pm} = \sqrt{2\left(1 \pm e^{-2 |\alpha_0|^2}\right)}.
\label{N_cat}
\eeq
The corresponding density operator is given by
\begin{align}
&\rho_{h} = \frac{1}{2}\left(\ketbra{cat_-}{cat_-}{C}\otimes\ketbra{0}{0}{D} + \ketbra{cat_+}{cat_+}{C}\otimes\ketbra{1}{1}{D} \right.\nonumber\\
&+ \left.\ketbra{cat_-}{cat_+}{C}\otimes\ketbra{0}{1}{D} + \ketbra{cat_+}{cat_-}{C}\otimes\ketbra{1}{0}{D}\right).
\label{hybrid_CD}
\end{align}

Due to the mathematical complexity of the hybrid entangled state in Eq.~(\ref{hybrid_ket}), especially when describing the teleportation of the CV mode in section~\ref{section_teleport_cv_mode}, we will find it useful to introduce approximations to such a hybrid entangled state in the cases of large and small cat states. The mathematical approximations we utilize will have high fidelity with Eq.~(\ref{hybrid_ket}) and  will allow for improved analytical insight.  Deviation from results produced using Eq.~(\ref{hybrid_ket}) directly will be inconsequential \cite{morin2014remote,huang2019engineering,nielson2007transforming}. Our main interest will be  large cat states, as they tend to be of wider interest in a range of quantum information protocols, such as information processing \cite{gilchrist2004}. However, we will still investigate the small cat-state limit, mainly because the results in this regime provide firm upper bounds on the teleportation fidelities as the cat states approach the so-called `kitten' states. 

\paragraph{Large cat states ($|\alpha_0|>1$)}
In order to give an approximation for large cat states, we first need to rearrange the terms in Eq.~(\ref{hybrid_ket}) by substituting
\beq
\ket{0} = \left(\ket{+}+\ket{-}\right)/\sqrt{2}, \; \ket{1} = \left(\ket{+}-\ket{-}\right)/\sqrt{2},
\label{pm_01}
\eeq
giving
\begin{align}
\ket{\psi}_{h} = \frac{1}{\sqrt{2}}&  \;\left( \;
\frac{\ket{cat_-}_C + \ket{cat_+}_C}{\sqrt{2}}  \ket{+}_D \right.\nonumber\\
& + \left. \frac{\ket{cat_-}_C - \ket{cat_+}_C}{\sqrt{2}}\ket{-}_D\right).
\end{align}
When $|\alpha_0|>1$, the states in mode $C$ can be approximated by coherent states, and our hybrid entangled state becomes \cite{morin2014remote}\cite{huang2019engineering}
\beq
\ket{\psi}_{h}^{(large)} \approx \frac{\ket{\alpha_0}_C\ket{+}_D - \ket{-\alpha_0}_C\ket{-}_D}{\sqrt{2}} \ .
\label{hybrid_large_alpha}
\eeq
The corresponding density matrix is
\begin{align}
&\rho_h^{(large)} \approx \nonumber\\
&\quad\frac{1}{2}(\ketbra{\;\alpha_0}{\;\alpha_0}{C} \otimes\ketbra{+}{+}{D}
 +\ketbra{-\alpha_0}{-\alpha_0}{C} \otimes \ketbra{-}{-}{D} \nonumber\\
&\quad -  \ketbra{\;\alpha_0}{-\alpha_0}{C} \otimes \ketbra{+}{-}{D}
 -\ketbra{-\alpha_0}{\;\alpha_0}{C} \otimes  \ketbra{-}{+}{D}).
\label{hybrid_large_alpha_rho}
\end{align}
Eqs.~(\ref{hybrid_large_alpha}) and (\ref{hybrid_large_alpha_rho}) will be useful for the calculation of the fidelity of CV-mode teleportation later in section~\ref{section_teleport_cv_mode}.

\paragraph{Small cat states ($|\alpha_0|<0.5$)}
We can approximate small cat states with the Single-Mode Squeezed Vacuum (SMSV) state and the Single-Photon Subtracted Squeezed Vacuum (1PS) state \cite{huang2019engineering}\cite{nielson2007transforming},
\begin{align}
\ket{cat_+} &\approx \ket{SMSV} = S(\zeta)\ket{0}, \\
\ket{cat_-} &\approx \ket{1PS} = \frac{\hat{a}S(\zeta)\ket{0}}{\sinh \zeta} = S(\zeta)\ket{1},
\label{cat_approximation}
\end{align}
where $S(\zeta)$ is the single-mode squeezing operator, and $\zeta = s e^{i\theta}$ with $s$ being the single-mode squeezing parameter.
For an arbitrary value of $\alpha_0$ in Eq.~(\ref{cat}), the fidelity between $\ket{cat_-}$ and $\ket{1PS}$ becomes maximized when $\alpha_0$ is real and \cite{nielson2007transforming}
\beq
s = \frac{\sqrt{9+4 \alpha_0^4} - 3}{2 \alpha_0^2}.
\label{s}
\eeq
Under these conditions, the fidelity between the SMSV (or 1PS) state and $\ket{cat_+}$ (or $\ket{cat_-}$) increases as $\alpha_0$ decreases, and the fidelity approaches unity as $\alpha_0$ approaches $0.5$ \cite{nielson2007transforming}.
For simplicity, we will henceforth set $\theta=0$ and write the single-mode squeezing operator as $S(s)$.
The original hybrid entangled state in Eq.~(\ref{hybrid_ket}) can therefore be approximated as \cite{huang2019engineering}
\beq
\ket{\psi}_h^{(small)} \approx  \frac{  S_C(s)\ket{1}_C \otimes \ket{0}_{D} +  S_C(s)\ket{0}_C\otimes \ket{1}_{D}}{\sqrt{2}}.
\label{approx_hybrid}
\eeq
The corresponding density matrix is given by
\begin{align}
\rho_h^{(small)} &\approx \frac{1}{2}\left[S(s)\ketbra{1}{1}{C} S(s)^\dagger \otimes \ketbra{0}{0}{D} \right.\nonumber\\
&\quad + S(s)\ketbra{0}{0}{C} S(s)^\dagger \otimes \ketbra{1}{1}{D} \nonumber\\
&\quad + S(s)\ketbra{1}{0}{C} S(s)^\dagger \otimes \ketbra{0}{1}{D} \nonumber\\
&\quad + \left. S(s)\ketbra{0}{1}{C} S(s)^\dagger\otimes\ketbra{1}{0}{D}\right].
\label{approx_hybrid_rho}
\end{align}
From now on, we will assume that $\alpha_0$ is real in all our calculations. For simplicity, we will also replace the approximation signs following $\rho_h^{(large)}$ and $\rho_h^{(small)}$ with equalities. Henceforth, whenever we mention large (small) cat states, or large (small) $\alpha_0$, we are refering to $\alpha_0>1$ ($\alpha_0<0.5$).

\paragraph{Intermediate cat states ($0.5 \le \alpha_0 \le 1$)}
In this case, we use the exact form of the  hybrid entangled state in Eq.~(\ref{hybrid_ket}) to carry out our calculations. These lead to relations that are perhaps  less illuminating as the approximate solutions above, and as such relegate this  exposition to appendix~\ref{appendix_cat_exact}. We shall see how the results from this exact solution match the approximate solutions in the relevant range.

Hybrid entanglement with large cat states can be deterministically generated by a weak and dispersive light-matter interaction \cite{vanloock2006hybrid}, or by a cross-Kerr nonlinear interaction between a coherent state and a photon-number qubit state \cite{gerry1999generation}\cite{jeong2005using}. However, due to experimental challenges in nonlinear optics, hybrid entanglement with small cat states can be more easily produced by using linear optics and a probabilistic heralded scheme
 \cite{morin2013remote}\cite{sychez2018entanglement}\cite{huang2019engineering}\cite{knill2001ascheme}.
The latter setup takes a small cat state  ($\ket{cat_+}$) as an input, and produces hybrid entanglement between small cat states and qubit states.
The advantage of the heralded scheme is that the loss in the heralding channel only affects the success rate but not the fidelity of the final entanglement. Such a scheme is particularly beneficial for long-distance communication \cite{huang2019engineering}. In addition, the heralded setup in \cite{sychez2018entanglement} has the potential to be extended for cat states with higher amplitudes.

\section{Direct distribution of a hybrid entangled state through lossy channels}
\label{direct_distribution}

The hybrid entangled state is given in Eq.~(\ref{hybrid_CD}), where the cat states have amplitude $\alpha_0$. 
We refer to the modes as given in Fig.~\ref{direct_setup}. That is, we denote the two modes of the entanglement as $A$ and $B$, where $A$ is the CV mode and $B$ is the DV mode. For each mode, the channel attenuation can be modeled by a beam splitter  with corresponding transmissivities $T_A$ and $T_B$. The photonic loss in \emph{either} the DV or CV mode has been studied in previous papers \cite{yin2019coherent,zhang2019improving,parker2017hybrid, parker2020photonic}. However, when the loss is present in \emph{both} the DV and CV mode, the resulting density matrix becomes complex and has not been hitherto determined. In this section, we calculate this density matrix and then determine the resulting fidelity of the hybrid state after direct distribution from the satellite.

At the first beam splitter, the CV mode $A$ is mixed with the auxiliary vacuum mode $\epsilon_A$ to give the output modes $A'$ and $\epsilon_{A'}$.
For a general bi-coherent input state $\ket{\alpha}_A\ket{\alpha_\epsilon}_{\epsilon_A}$, the beam splitter transformation gives  
\begin{align}
&\ket{\alpha}_A\ket{\alpha_\epsilon}_{\epsilon_A} \rightarrow \nonumber\\
&\ket{\sqrt{T_A}\alpha - \sqrt{1-T_A}\alpha_\epsilon}_{A'} \ket{\sqrt{T_A}\alpha_\epsilon + \sqrt{1-T_A}\alpha}_{\epsilon_{A'}}.
\label{BStransformation_CV}
\end{align}
At the second beam splitter, the DV mode $B$ is mixed with the auxiliary vacuum mode $\epsilon_B$, giving the output modes $ B'$ and $\epsilon_{B'}$.
Let the creation operators of the four modes be $\hat{a}_B^{\dagger}$, $\hat{a}_{\epsilon_B}^{\dagger}$,$\hat{a}_{B'}^{\dagger}$ and $\hat{a}_\eb^{\dagger}$, the beam splitter transformation can then be written as
\begin{align}
\hat{a}^{\dagger}_B&\rightarrow \sqrt{T_B}\,\hat{a}^{\dagger}_{B'} + \sqrt{1-T_B}\,\hat{a}^{\dagger}_{\epsilon_{B'}},\label{BStransformation_DV}\\
\hat{a}^{\dagger}_{\epsilon_B}&\rightarrow \sqrt{T_B}\,\hat{a}^{\dagger}_{\epsilon_{B'}} - \sqrt{1-T_B}\,\hat{a}^{\dagger}_{B'}.
\label{BStransformation_DV_au}
\end{align}
After applying the beam splitter transformations, the density matrix of the hybrid entangled state (Eq.~(\ref{hybrid_CD})) becomes $\rho_{A^{'}\epsilon_{A'} B^{'} \epsilon_{B'}}$. The transmitted state can be found by tracing out the auxiliary output modes $\epsilon_{A'}$ and $\epsilon_{B'}$. Mode $\epsilon_{B'}$ can be easily traced out, since it is a DV mode. The CV mode $\epsilon_{A'}$ can be traced out by applying the integration
\beq
\int \frac{d^2 \beta}{\pi} \bra{\beta}\rho_{A^{'}\epsilon_{A'} B^{'} \epsilon_{B'}}\ket{\beta}_{\epsilon_{A'}},
\eeq
where $\beta$ is a generic complex number.
This integration can be calculated by applying the following well-known identities for a general coherent state $\ket{\alpha}_{\epsilon_{A'}}$
\begin{align}
\int \frac{d^2 \beta}{\pi}\braket{\beta}{\pm \alpha}_{\epsilon_{A'}}\braket{\pm \alpha}{\beta}_{\epsilon_{A'}} &= 1,\nonumber\\
\int \frac{d^2 \beta}{\pi}\braket{\beta}{\pm \alpha}_{\epsilon_{A'}}\braket{\mp \alpha}{\beta}_{\epsilon_{A'}} &= \exp\left(-2 |\alpha|^2\right).
\label{trace_cv}
\end{align}
After tracing out the auxiliary modes (where the details can be found in appendix~\ref{appendix_direct_distribution}),
we find the directly-distributed state
\begin{align}
&\rho_{A'B'}=\tr_{\epsilon_{A'} \epsilon_{B'}}(\rho_{A^{'}\epsilon_{A'} B^{'} \epsilon_{B'}})=
\frac{1}{2}\left\{\ketbra{\sqrt{T_A}\alpha_0}{\sqrt{T_A}\alpha_0}{A}\right.\nonumber\\
&\otimes\left[(a_1 + a_2)\ketbra{0}{0}{B}+ a_3 \ketbra{1}{1}{B} + a_4\ketbra{0}{1}{B} + a_4 \ketbra{1}{0}{B}\right]\nonumber\\
&+\ketbra{-\sqrt{T_A}\alpha_0}{-\sqrt{T_A}\alpha_0}{A}\nonumber\\
&\otimes\left[(a_1 + a_2)\ketbra{0}{0}{B}+ a_3 \ketbra{1}{1}{B} - a_4\ketbra{0}{1}{B} - a_4 \ketbra{1}{0}{B}\right]\nonumber\\
&+ e^{-2(1-T_A)\alpha_0^2}\ketbra{\sqrt{T_A}\alpha_0}{-\sqrt{T_A}\alpha_0}{A}\nonumber\\
&\otimes\left[(a_1 - a_2)\ketbra{0}{0}{B}+ a_3 \ketbra{1}{1}{B} + a_4\ketbra{0}{1}{B} - a_4 \ketbra{1}{0}{B}\right]\nonumber\\
&+ e^{-2(1-T_A)\alpha_0^2}\ketbra{-\sqrt{T_A}\alpha_0}{\sqrt{T_A}\alpha_0}{A}\nonumber\\
&\left.\otimes \left[(a_1 - a_2)\ketbra{0}{0}{B}+ a_3 \ketbra{1}{1}{B} - a_4\ketbra{0}{1}{B} + a_4 \ketbra{1}{0}{B}\right]\vphantom{\sqrt{T_A}}\right\},
\label{arbitrary_alpha}
\end{align}
where
\beq
a_1 = \frac{1-T_B}{N_+^2},\;
a_2 = \frac{1}{N_-^2}, \;
a_3= \frac{T_B}{N_+^2},\;
a_4 = \frac{\sqrt{T_B}}{N_+ N_-},
\eeq
with $N_\pm$ defined in Eq.~(\ref{N_cat}).

After calculating the directly-distributed state, we will compare it to the original hybrid entangled state - using fidelity as our key performance metric.
However, since in some applications  the logarithmic negativity of the entangled state is useful, that metric will also be discussed.
In the density matrix formalism, the fidelity between the perfect hybrid entangled state ($\rho_h$ in Eq.~(\ref{hybrid_CD})) and an arbitrary state $\rho$ is given by \cite{jozsa1994fidelity}
\beq
F = \left[\tr\left( \sqrt{ \sqrt{\rho_h} \, \rho \, \sqrt{\rho_h}} \right)\right]^2.
\label{fidelity_rho}
\eeq
As the fidelity approaches one, the directly-distributed state becomes identical to the hybrid state created by the satellite (we will assume the hybrid state is initially perfect when produced anywhere).
The negativity of an arbitrary state $\rho$ is defined by \cite{vidal2002computable}
\beq
E_N(\rho) = \frac{1}{2}\sum_i\left( |\lambda_i|-\lambda_i \right),
\eeq
where $\lambda_i$ are the eigenvalues of the partial transpose of the entangled state $\rho$.
The logarithmic negativity ($0 \leq E_{LN} \leq 1$) is related to the negativity by \cite{vidal2002computable}
\beq
E_{LN}(\rho) = \log_2[1+2 E_N(\rho)].
\label{ELN_rho}
\eeq
When $E_{LN}>0$, $\rho$ is an entangled state.

\section{Teleportation through a TMSV channel}
\label{teleportation}

In this section, we detail the teleportation of an input mode of the hybrid entangled state  through the attenuated TMSV teleportation channel (Fig.~\ref{teleported_setup}). The CV entangled state $A-B$ is created on the satellite, then transmitted down to Alice and Bob, with channel transmissivities $T_A$ and $T_B$, respectively, resulting in the attenuated TMSV channel $A'-B'$. When the input mode is the DV mode (mode $D$) of the hybrid entangled pair $C-D$, the CV teleportation channel teleports $D$ to $B''$, resulting in the entanglement between $C$ and $B''$. Otherwise, when the input mode is the CV mode, $C$ is teleported to $B''$, resulting in the entanglement between $D$ and $B''$.

Experimentally, the CV entanglement ($A-B$) can be produced by letting two single-mode squeezed vacuum states interfere through a balanced beam splitter.\footnote{In this work, we do not take into account the loss at this beam splitter. However, such additional losses can be readily mapped to a re-scaling in both the initial squeezing parameter and channel loss  \cite{takeda2013gaintuning}.} Theoretically, the CV entanglement can be created by applying a two-mode squeezing operator on a two-mode vacuum state to obtain a TMSV state
\beq
\ket{TMSV} = S(\xi)\ket{0,0} = \exp\left(\xi \hat{a}_A^\dagger \hat{a}_B^\dagger - \xi^* \hat{a}_A\hat{a}_B\right)\ket{0,0},
\label{TMSV_ket}
\eeq
where $\xi = r e^{i\phi}$, with $r$ being the two-mode squeezing parameter (henceforth referred to as the initial squeezing), and $\phi$ being the phase. $\hat{a}_l$ and $\hat{a}_l^\dagger$ represent the annihilation and creation operators of the optical field, respectively, with $l \in \{A,B\}$ denoting the spatial modes.
Let $\hbar =1 / 2 $, CV entanglement is encoded in the dimensionless position and momentum quadrature operators of the optical field,
$\hat{x_l} = (\hat{a_l}^\dagger+\hat{a_l})/{2}$ and $\hat{p_l}=i(\hat{a_l}^\dagger-\hat{a_l})/{2}$,
respectively, where $i$ denotes the imaginary unit. In the following two subsections, we will present the mathematical models that describe the teleportation of either the DV or CV mode of a hybrid entangled state through an attenuated TMSV channel.\footnote{The models developed in this work have been verified by teleporting a simple vacuum state through a symmetric channel, which produces the same fidelity as the analytical results in \cite{takeoka2003continuous} and \cite{lie2019limitations}.}

\subsection{Teleporting the DV mode of hybrid entanglement}
\label{section_dv_teleportation}
In order to describe the teleportation of the DV mode of the hybrid entanglement through the CV teleportation channel, we will use the Wigner function representation. We first study a general input state with the Wigner function $W_{in}(\alpha)$, where $\alpha = x + i p$  with non-zero real values $x$ and $p$.
Let $G_{\sigma}(\alpha)$ denote a Gaussian function, with $x$ and $p$ denoting two uncorrelated random variables with equal variance $\sigma$
\beq
G_\sigma(\alpha) =\frac{1}{\pi\sigma} \exp\left(-\frac{|\alpha|^2}{\sigma}\right)=\frac{1}{\pi\sigma} \exp\left(-\frac{x^2 + p^2}{\sigma}\right).
\label{Gaussian}
\eeq
From \cite{takeda2013gaintuning}, the teleported Wigner function after the use of the TMSV channel is
\begin{align}
W_{tel}(\alpha_{B''}) &= \frac{1}{g^2}\left[W_{in}*G_{\sigma}\right]\left(\frac{\alpha_{B''}}{g}\right)\nonumber\\
&= \frac{1}{g^2}\iint dx dp W_{in}(\alpha) G_{\sigma}\left(\frac{\alpha_{B''}}{g}-\alpha\right).
\label{general_teleportation}
\end{align}
Here $G_{\sigma}$ is the Gaussian function defined in Eq.~(\ref{Gaussian}) with variance $\sigma$ given by
\begin{align}
\sigma & = \frac{1}{4g^2}\;\,\left[\;\, e^{+2r}\left(g\sqrt{T_A}-\sqrt{T_B}\right)^2 \right.\nonumber\\
&\qquad\quad+e^{-2r}\left(g\sqrt{T_A}+\sqrt{T_B}\right)^2   \nonumber\\
&\qquad\quad + 2g^2 (1-T_A) + 2(1-T_B) \left.\vphantom{\left(\sqrt{T_A}\right)^2}\right],
\label{sigma_general}
\end{align}
where $g$ is the teleportation gain. For fixed values of $T_A$ and $T_B$, the calculation of the logarithmic negativity has shown that the optimal gain $g$ increases with $r$, and reaches its maximal value at $r\simeq 2.5$ \cite{do2019hybrid}. From now on, for simplicity, whenever we mention the 'optimal gain', we are referring to the optimal gain when the initial squeezing satisfies $r>2.5$. When $T_A = T_B=1$, the optimal gain is approximately $g\approx 1$, and we have $\sigma= \exp(-2 r)$ \cite{braustein1998teleportationCV}. When the channels are asymmetric, i.e., $T_A\neq T_B$, the teleportation (in terms of logarithmic negativity and fidelity) is optimized when $g \approx \sqrt{\frac{T_B}{T_A}}$ \cite{do2019hybrid}. Note that tuning the gain further does not improve the fidelity much, thus, from now on, we will simply replace the approximation signs by the equal signs.

When the input mode is the DV mode ($D$) of a hybrid entangled pair $C-D$, the subspace of $C$ stays the same, while the subspace of $D$ is transformed.
Mode $D$ can be decomposed to different terms $\ketbra{m}{n}{D}$ ($m,n \in \{0,1\}$), with Wigner functions  $W^{\ketbra{m}{n}{D}}(\alpha_D)$.
The teleportation of each term is described by Eqs.~(\ref{Gaussian}), (\ref{general_teleportation}) and (\ref{sigma_general}), leading to
\beq
W^{\ketbra{m}{n}{D}}(\alpha_D) \rightarrow \frac{1}{g^2}\left[W_{in}^{\ketbra{m}{n}{D}}(\alpha_D)*G_{\sigma}\right]\left(\frac{\alpha_{B''}}{g}\right).
\label{Fock_teleportation}
\eeq
The teleportation transformation has been previously calculated for the case where mode $D$ is part of a Bell state  \cite{takeda2013gaintuning}\cite{do2019hybrid}.
When mode $D$ is part of hybrid entanglement, we recognize that the initial hybrid entanglement $\ket{\psi}_h$ (Eq.~(\ref{hybrid_CD})) has a form that is similar to a Bell state, where $\ket{cat_-}_C$ corresponds to $\ket{1}_C$, and $\ket{cat_+}_C$ corresponds to $\ket{0}_C$. By transforming the subspace of mode $D$ similar to that described in \cite{takeda2013gaintuning}\cite{do2019hybrid}, and then converting the Wigner formalism to the density-matrix formalism, we find the DV-mode teleported state
\begin{align}
\rho_{tel} &= \sum_{k=-1}^{\infty}\rho_k, \,\textrm{where} \nonumber\\
\hat{\rho}_{k} &= a_{k} \ketbra{cat_+}{cat_+}{C}\otimes\ketbra{k}{k}{B''} \nonumber \\
&+b_{k}\ketbra{cat_-}{cat_+}{C}\otimes\ketbra{k}{k+1}{B''} \nonumber\\
&+ b_{k} \ketbra{cat_+}{cat_-}{C}\otimes\ketbra{k+1}{k}{B''}\nonumber\\
 &+c_{k} \ketbra{cat_-}{cat_-}{C}\otimes\ketbra{k+1}{k+1}{B''} ,
\label{teleported_output}
\end{align}
where $a_k$, $b_k$ and $c_k$  are defined by
\begin{align}
	a_k &= \frac{1}{2}T_{1,1\rightarrow k,k} \,( k\geq 0),\text{or} \;0  \; (k = -1),\nonumber\\
		b_k & = \frac{1}{2}T_{1,0\rightarrow k+1, \,k}\, (k\geq 0), \text{or} \;0 \;(k=-1) ,\nonumber\\
		c_k & = \frac{1}{2}T_{0,0\rightarrow k+1, \, k+1} \,(k\geq -1), \,\textrm{and where}
		\label{abc}
\end{align}
\begin{align}
T_{0,0\rightarrow k,k} &= \frac{2(\gamma-1)^k}{(\gamma+1)^{k+1}},\nonumber\\
T_{1,1\rightarrow k,k} &= \frac{2(\gamma-1)^{k-1}}{(\gamma+1)^{k+2}}\left[(\gamma-2g^2+1)(\gamma-1)+4kg^2\right]\nonumber\\
T_{1,0\rightarrow k+1,\,k} &= \frac{4g\sqrt{k+1}(\gamma-1)^k}{(\gamma+1)^{k+2}},
\label{T_transform}
\end{align}
with $\gamma \equiv g^2(2\sigma + 1)$.

In the density matrix representation, the teleported state $\rho_{tel}$ can be compared to the perfect hybrid entangled state $\rho_h$ by using the measures of fidelity (Eq.~(\ref{fidelity_rho})) and  logarithmic negativity (Eq.~(\ref{ELN_rho})). The simulation results are shown in section~\ref{result}.

\subsection{Teleporting the CV mode of hybrid entanglement}
\label{section_teleport_cv_mode}
In this section, we will discuss how to teleport the CV mode $C$ of the hybrid entangled state $D-C$ through the TMSV channel to obtain the final entangled state $D-B''$. (Note that, in this section, we swap the order of modes $C$ and $D$ in Fig.~\ref{teleported_setup}). Due to mathematical complexity, we will rewrite all required states in the characteristic-function formalism and approximate the hybrid entangled state ($D-C$) for the limits of large and small cat states in section~\ref{hybrid_entanglement_math}.


\subsubsection{The characteristic function}
\label{section_define_cf}
For an arbitrary state $\rho$, the usual characteristic function is defined as 
\beq
\bigchi_{\rho}(\beta) = \tr\left[ \rho D(\beta)\right],
\label{characteristic_function_define}
\eeq
with complex number $\beta=x + ip$ and displacement operator
$
D(\beta) = e^{-|\beta|^2/2}e^{\beta\hat{a}^\dagger}e^{-\beta^* \hat{a}}.
$ 
For example, for a vacuum state $\ket{0}$, the characteristic function is given by
\beq
\bigchi_{vac}(\beta) = \exp{\left[ -|\beta|^2/2 \right]}.
\label{vac_chi}
\eeq
For a single-mode squeezed vacuum (SMSV) state,  let $S(\zeta)$ be the single-mode squeezing operator with $\zeta = s e^{i\theta}$, where $s$ is the single-mode squeezing parameter (which is different from the two-mode squeezing parameter $r$ in Eq.~(\ref{TMSV_ket})) and $\theta$ is the phase. The characteristic function of the SMSV state can be written as \cite{farias2009thesis} \cite{seshadreesan2015nongaussian}
\begin{align}
\bigchi_{SMSV}(\beta) &= \exp\left( -\frac{1}{2}|\beta \cosh s - e^{i\theta}\beta^* \sinh s|^2 \right)\\
&=\exp\left[-\frac{1}{2}(x^2e^{-2s}+ p^2e^{2s}) \right],
\label{SMSV}
\end{align}
where the last equality has assumed  $\theta=0$.
For a TMSV state given by Eq.~(\ref{TMSV_ket}), when there is no channel loss, the characteristic function is given by
\cite{seshadreesan2015nongaussian},
\begin{align}
&\bigchi_{TMSV}(\beta_A,\beta_B) = \exp \left[ -\frac{1}{2} \left( |\beta_A \cosh r + \beta_B^* e^{i\phi}\sinh r|^2 \right.\right.\nonumber\\
&\qquad\qquad\qquad\qquad\left.\left. + |\beta_B \cosh r + \beta_A^*e^{i\phi}\sinh r|^2 \right) \vphantom{\frac{1}{2}}\right]\\
&=\exp \left\{ -\frac{e^{2r}}{4}\left[\left(x_A-x_B\right)^2 + \left( p_A+p_B\right)^2 \right]\right.\nonumber\\
& \qquad\quad\left. -\frac{e^{-2r}}{4}\left[\left(x_A+x_B\right)^2 + \left( p_A-p_B\right)^2 \right]\right\},
\label{TMSV_chi}
\end{align}
where the last equality has used $\phi=\pi$ for optimal teleportation fidelity.

For a tensor product of arbitrary density matrices, the characteristic function is a product of individual characteristic functions. For example, for an arbitrary separable state $\rho=\rho_1\otimes\rho_2$, we have \cite{dellanno2018nongaussian}
\beq
\bigchi_\rho(\beta_1,\beta_2) = \tr\left[\rho D(\beta_1)D(\beta_2)\right] = \bigchi_{\rho_1}(\beta_1)\bigchi_{\rho_2}(\beta_2) .
\label{characteristic_terms}
\eeq
The characteristic function is also linear with respect to the density operator, as can be seen from the linearity of the trace function in Eq.~(\ref{characteristic_function_define}).

\subsubsection{Characteristic function of an attenuated TMSV state}
Again, we model the two down-link channels by two beam splitters with transmissivities $T_A$ and $T_B$, respectively. At the beam splitters, modes $A$ and $B$ are mixed with the auxiliary vacuum modes $\epsilon_A$ and $\epsilon_B$, respectively, giving the output modes $\{A',\epsilon_{A'}\}$, and $\{ B', \epsilon_{B'}\}$, respectively. Let $t_l = \sqrt{T_l}, r_l = \sqrt{1-T_l}$ with $l \in \{A,B\}$, 
the beam splitter transformation can be given by
\begin{align}
\beta_{l'} &= t_l \beta_l - r_l \beta_{\epsilon_l}, \quad \beta_{\epsilon_{l'}} = r_l \beta_l + t_l \beta_{\epsilon_l}.
\label{bs_cv}
\end{align}
The four-mode characteristic equation before the beam splitter is given by multiplying the characteristic functions of the TMSV state with the vacuum states
\begin{align}
\bigchi_{TMSV}(\beta_A,\beta_B)\bigchi_{vac}(\beta_{\epsilon_A})\bigchi_{vac}(\beta_{\epsilon_B}).
\end{align}
The beam splitter transformation can be performed by substituting
$\beta_l = \;t_l \beta_{l'} + r_l \beta_{\epsilon_{l'}}$ and $\beta_{\epsilon_l} = - r_l \beta_{l'} + t_l \beta_{\epsilon_{l'}}$
 into the above equation. It has been shown that the tracing out the vacuum modes is equivalent to setting $\beta_{\epsilon_{l'}}=0$ \cite{dellanno2010realistic}. The resulting attenuated TMSV state is given by
\begin{align}
&\bigchi_{TMSV}^{(T_A,T_B)}(\beta_{A'},\beta_{B'}) = \bigchi_{TMSV}(\sqrt{T_A}\beta_{A'},\sqrt{T_B}\beta_{B'})\nonumber\\
&\qquad\times\bigchi_{vac}(-\sqrt{1-T_A}\beta_{A}')\bigchi_{vac}(-\sqrt{1-T_B}\beta_{B}').
\label{TMSV_tatb}
\end{align}

\subsubsection{Characteristic function of a hybrid entangled state}
\label{approximate_hybrid}
The exact density matrix of the hybrid entangled state is given by Eq.~(\ref{hybrid_CD}), in which the cat states $\ket{cat_\pm}$  have the characteristic functions \cite{Girish2013QuantumOptics}
\begin{align}
\bigchi_{cat_\pm}(\beta) = &\frac{1}{N_{\pm}^2}\left\{2 e^{-\frac{|\beta|^2}{2}} \cos\left[ 2\textrm{Im}\left( \beta\alpha_0^* \right)\right]\right.\nonumber\\
&\left. \pm e^{-\frac{1}{2}|\beta+2\alpha_0|^2} \pm e^{-\frac{1}{2}|\beta-2\alpha_0|^2}\right\}.
\end{align}
However, due to the mathematical complexity, especially when calculating the matrix $\ketbra{cat_\pm}{cat_\mp}{}$ in Eq.~(\ref{hybrid_CD}), we will approximate the hybrid entangled state for the cases of large $\alpha_0$ (Eq.~(\ref{hybrid_large_alpha_rho})) and small $\alpha_0$ (Eq.~(\ref{approx_hybrid_rho})). For hybrid states with intermediate sizes, an exact calculation can be found in appendix~\ref{appendix_cat_exact}.

\paragraph {When the cat state is large} the hybrid entangled state can be approximated by $\rho_h^{(large)}$ in Eq.~(\ref{hybrid_large_alpha_rho}).
Given the definition of the characteristic function (Eq.~(\ref{characteristic_function_define})), and the form of the density matrix of the approximate  hybrid state (in Eq.~(\ref{hybrid_large_alpha_rho})), we will find it useful to introduce a new function for a general matrix $M$ 
\beq
X_{M}(\beta) = \tr\left[M D(\beta) \right].
\label{X_function}
\eeq
From the linearity of the trace function, the above function is also linear with respect to the matrix $M$ \cite{Girish2013QuantumOptics}. For a tensor product of the form $M = M_1 \otimes M_2$, we have
\begin{align}
X_M (\beta_1,\beta_2)&= \tr\left[M_1 \otimes M_2 \; D_1(\beta_1)\otimes D
_2(\beta_2)\right] \nonumber\\
&= X_{M_1}(\beta_1)X_{M_2}(\beta_2),
\label{X_tensor}
\end{align}
where the last equality makes use of the tensor product property of the trace function.
From the linearity and tensor product property above, the characteristic function of $\rho_h^{(large)}$ can thus be found by applying Eq.~(\ref{X_tensor}) to a modified  Eq.~(\ref{hybrid_large_alpha_rho}) where the order of modes $C$ and $D$ is swapped (see discussion start of section~\ref{section_teleport_cv_mode}), giving
\begin{align}
 \bigchi_{h}^{(large)}(\beta_D,\beta_C)
= &\frac{1}{2}\left[X_{\ketbra{+}{+}{}}(\beta_D) X_{\ketbra{\alpha_0}{\alpha_0}{}}(\beta_C) \right.\nonumber\\
&+X_{\ketbra{-}{-}{}}(\beta_D) X_{\ketbra{-\alpha_0}{-\alpha_0}{}}( \beta_C) \nonumber\\
&-X_{\ketbra{+}{-}{}}(\beta_D) X_{\ketbra{\alpha_0}{-\alpha_0}{}}(\beta_C) \nonumber\\
&-\left. X_{\ketbra{-}{+}{}}(\beta_D) X_{\ketbra{-\alpha_0}{\alpha_0}{}}(\beta_C)\right].
\label{hybrid_chi_large}
\end{align}
In the following, we will explicitly calculate the individual terms in the above equation.

From the change of basis in Eq.~(\ref{pm_01}), we can show
\begin{align}
&X_{\ketbra{+}{+}{}}(\beta) \nonumber\\
&\quad =  \frac{1}{2}\left[X_{\ketbra{0}{0}{}}(\beta) + X_{\ketbra{1}{1}{}}(\beta) +  X_{\ketbra{0}{1}{}}(\beta) + X_{\ketbra{1}{0}{}}(\beta)\right],\nonumber\\
&X_{\ketbra{-}{-}{}}(\beta) \nonumber\\
&\quad =   \frac{1}{2}\left[X_{\ketbra{0}{0}{}}(\beta) + X_{\ketbra{1}{1}{}}(\beta) -  X_{\ketbra{0}{1}{}}(\beta) - X_{\ketbra{1}{0}{}}(\beta)\right],\nonumber\\
&X_{\ketbra{+}{-}{}}(\beta) \nonumber\\
&\quad =   \frac{1}{2}\left[X_{\ketbra{0}{0}{}}(\beta) - X_{\ketbra{1}{1}{}}(\beta) -  X_{\ketbra{0}{1}{}}(\beta) + X_{\ketbra{1}{0}{}}(\beta)\right],\nonumber \\
&X_{\ketbra{-}{+}{}}(\beta) \nonumber\\
&\quad =   \frac{1}{2}\left[X_{\ketbra{0}{0}{}}(\beta) - X_{\ketbra{1}{1}{}}(\beta) +  X_{\ketbra{0}{1}{}}(\beta) - X_{\ketbra{1}{0}{}}(\beta)\right],
\label{cf_pm}
\end{align}
where 
\begin{align}
X_{\ketbra{0}{0}{}}(\beta) &= \bra{0}D(\beta)\ket{0}=e^{-|{\beta}|^2/2},\nonumber\\
X_{\ketbra{1}{1}{}}(\beta) &= \bra{1}D(\beta)\ket{1} = e^{-|{\beta}|^2/2}(1-|{\beta}|^2),\nonumber\\
X_{\ketbra{0}{1}{}}(\beta) &= \bra{1}D(\beta)\ket{0}=\beta e^{-|{\beta}|^2/2 }, \nonumber\\
X_{\ketbra{1}{0}{}}(\beta) &= \bra{0}D(\beta)\ket{1}= \left[\bra{0}D^\dagger(-\beta)\right]\ket{1}= -{\beta}^* e^{-|{\beta}|^2/2}.
\label{cf_fock}
\end{align}
In Eq.~(\ref{cf_fock}) we have used the appropriate relations for the displaced number states \cite{glauber1963coherent,sudarshan1963equivalence, deoliveira1990properties}.
For the terms $\ketbra{\pm\alpha}{\pm \alpha}{}$ and $\ketbra{\pm\alpha}{\mp \alpha}{}$ in Eq.~(\ref{hybrid_large_alpha_rho}), we have \cite{Girish2013QuantumOptics}
\begin{align}
 X_{\ketbra{\pm\alpha_0}{\pm\alpha_0}{}} (\beta)&=e^{-\frac{|\beta|^2}{2}}e^{\pm i 2 \textrm{Im}[\beta\alpha_0^*]}, \nonumber\\
 X_{\ketbra{\pm\alpha_0}{\mp\alpha_0}{}}(\beta) &= e^{-\frac{|\beta\pm 2\ao|^2}{2}}.
\label{cf_coh}
\end{align}
By substituting Eqs.~(\ref{cf_pm}), (\ref{cf_fock}) and (\ref{cf_coh}) into Eq.~(\ref{hybrid_chi_large}), we can find the characteristic function of the hybrid entangled state with large $\alpha_0$.

\paragraph {When the cat state is small} the hybrid entangled state can be approximated by $\rho_h^{(small)}$. By applying Eqs.~(\ref{characteristic_function_define}) and (\ref{X_tensor}) to Eq.~(\ref{approx_hybrid_rho}), the characteristic function of $\rho_h^{(small)}$ is given by
\begin{align}
\bigchi_{h}^{(small)}(\beta_D,\beta_C)
&= \frac{1}{2}\left[X_{\ketbra{0}{0}{}}(\beta_D) X_{S(s)\ketbra{1}{1}{}S(s)^\dagger}(\beta_C) \right.\nonumber\\
&\quad\;+ X_{\ketbra{1}{1}{}}( \beta_D) X_{S(s)\ketbra{0}{0}{}S(s)^\dagger}( \beta_C) \nonumber\\
&\quad\;+X_{\ketbra{0}{1}{}}(\beta_D) X_{S(s)\ketbra{1}{0}{}S(s)^\dagger}(\beta_C)\nonumber\\
&\quad \; + \left. X_{\ketbra{1}{0}{}}(\beta_D) X_{S(s)\ketbra{0}{1}{}S(s)^\dagger}(\beta_C)\right].
\end{align}
For the matrix $S(s)\ketbra{m}{n}{} S(s)^\dagger$ with $m,n \in \{0,1\}$ and $s$ defined in Eq.~(\ref{s}), $ X_{S(s)\ketbra{m}{n}{} S(s)^\dagger}(\beta)$
can be found by applying the relation
\beq
S^\dagger(s)D(\beta)S(s) = D(\tilde{\beta}),\,\textrm{with} \; \tilde{\beta} = \beta \cosh s - \beta^* \sinh s.
\eeq
As a result,
$X_{S(s)\ketbra{0}{0}{}S(s)^\dagger}(\beta)$ and $X_{S(s)\ketbra{1}{1}{}S(s)^\dagger}(\beta)$ can be found by substituting $\tilde{\beta}$ for $\beta$ in Eq.~(\ref{cf_fock}), giving \cite {farias2009thesis}\cite{seshadreesan2015nongaussian}
\begin{align}
X_{S(s)\ketbra{0}{0}{}S(s)^\dagger}(\beta) &=X_{\ketbra{0}{0}{}}(\tilde{\beta}),\\
X_{S(s)\ketbra{1}{1}{}S(s)^\dagger}(\beta) &=X_{\ketbra{1}{1}{}}(\tilde{\beta}).
\end{align}
Our calculation shows that $X_{S(s)\ketbra{0}{1}{}S(s)^\dagger}(\beta)$ and $S(s)X_{\ketbra{1}{0}{}S(s)^\dagger}(\beta)$ can also be found by the same substitution %
\begin{align}
X_{S(s)\ketbra{0}{1}{}S(s)^\dagger}( \beta) &= X_{\ketbra{0}{1}{}}(\tilde{\beta}),\\
X_{S(s)\ketbra{1}{0}{}S(s)^\dagger}(\beta) &= X_{\ketbra{1}{0}{}}(\tilde{\beta}).
\end{align}
(Note that, when $s=0$, $\tilde{\beta}$ becomes $\beta$, we  have $ X_{S(0)\ketbra{m}{n}{} S(0)^\dagger}(\beta) = X_{\ketbra{m}{n}{}}(\beta)$).
We can now write
\begin{align}
&\bigchi_{h}^{(small)}(\beta_D,\beta_C) = \nonumber\\
&\frac{1}{2}\left[X_{\ketbra{0}{0}{}}(\beta_D) X_{\ketbra{1}{1}{}}(\tilde{\beta}_C) \right.+ X_{\ketbra{1}{1}{}}( \beta_D) X_{\ketbra{0}{0}{}}(\tilde{ \beta}_C) \nonumber\\
&\quad+X_{\ketbra{0}{1}{}}(\beta_D) X_{\ketbra{1}{0}{}}(\tilde{\beta}_C)+ \left. X_{\ketbra{1}{0}{}}(\beta_D) X_{\ketbra{0}{1}{}}(\tilde{\beta}_C)\right].
\label{hybrid_chi}
\end{align}
By substituting Eq.~(\ref{cf_fock}) into the above equation, we can find the characteristic function of a hybrid entangled state with small $\alpha_0$.

\subsection{Fidelity}
\label{section_fidelity_cv}
In general, fidelities above the classical teleportation limit are considered useful in quantum communications (this is $\frac{2}{3}$ for a qubit state  and $\frac{1}{2}$ for a coherent state \cite{massar1995optimal})
The fidelity between two arbitrary two-mode states, for example, between the original hybrid entangled state $\bigchi_h(\beta_D,\beta_C)$  and the CV-mode teleported state $\bigchi_{tel}(\beta_D,\beta_{B''})$, is given by
\begin{align}
F &= \frac{1}{\pi^2}\iint d^2\beta_D d^2 \beta_{B''} \bigchi_h(\beta_D,\beta_{B''})\bigchi_{tel}(-\beta_D,-\beta_{B''}).
\label{eq_fidelity}
\end{align}
In order to verify our calculations, we check that the fidelity of a state with itself is always unity. For example, in the limit where the TMSV channel has no loss and has infinite squeezing, $\bigchi_{tel}(\beta_D,\beta_{B''})$ becomes identical to $\bigchi_h(\beta_D,\beta_{B''})$. For the case of large~$\alpha_0$, $\bigchi_h^{(large)}(\beta_D,\beta_{B''})$ is given by Eq.~(\ref{hybrid_chi_large}). The multiplication $\bigchi_h^{(large)}(\beta_D,\beta_{B''})\bigchi_h^{(large)}(-\beta_D,-\beta_{B''})$ creates a summation of 16 different terms. The summation can be simplified by exploiting the symmetry of the DV mode, which gives
\begin{align}
\int d^2\beta_D   X_{\ketbra{+}{+}{}}(\beta_D)X_{\ketbra{+}{+}{}}(-\beta_D)& \nonumber\\
\qquad=\int d^2\beta_D  \lvert X_{\ketbra{+}{+}{}}(\beta_D)\rvert ^2 &=\pi,\nonumber\\
\int d^2\beta_D   X_{\ketbra{-}{-}{}}(\beta_D)X_{\ketbra{-}{-}{}}(-\beta_D)&\nonumber\\
\qquad =\int d^2\beta_D   \lvert X_{\ketbra{-}{-}{}}(\beta_D)\rvert ^2 &=\pi,\nonumber\\
\int d^2\beta_D   X_{\ketbra{+}{-}{}}(\beta_D)X_{\ketbra{-}{+}{}}(-\beta_D)&=\pi,\nonumber\\
\int d^2\beta_D   X_{\ketbra{-}{+}{}}(\beta_D)X_{\ketbra{+}{-}{}}(-\beta_D)&=\pi,
\label{symmetrya}
\end{align}
while the other twelve integrations give zero, for example,
\begin{align}
&\int d^2\beta_D  X_{\ketbra{+}{+}{}}(\beta_D)X_{\ketbra{-}{-}{}}(-\beta_D)=0,\nonumber\\
&\int d^2\beta_D  X_{\ketbra{+}{+}{}}(\beta_D)X_{\ketbra{+}{-}{}}(-\beta_D)=0,\cdots
\label{symmetryb}
\end{align}
The fidelity of a large hybrid entangled state with itself is therefore given by the four non-zero terms
\begin{align}
&F_{self}^{(large)} = \nonumber\\
&\frac{1}{4}\left[ \frac{1}{\pi}\int d^2\beta_D|X_{\ketbra{+}{+}{}}(\beta_D)|^2 \right.\nonumber\\
&\qquad\times \frac{1}{\pi}\int d^2\beta_{B''}|X_{\ketbra{\alpha_0}{\alpha_0}{}}(\beta_{B''})|^2 \nonumber\\
&\;+\frac{1}{\pi}\int d^2\beta_D|X_{\ketbra{-}{-}{}}(\beta_D)|^2 \nonumber\\
&\qquad\times \frac{1}{\pi}\int d^2\beta_{B''}|X_{\ketbra{-\alpha_0}{-\alpha_0}{}}(\beta_{B''})|^2\nonumber\\
&\;+\frac{1}{\pi}\int d^2\beta_D X_{\ketbra{+}{-}{}}(\beta_D)X_{\ketbra{-}{+}{}}(-\beta_D) \nonumber\\
&\qquad\times \frac{1}{\pi}\int d^2\beta_{B''} X_{\ketbra{\alpha_0}{-\alpha_0}{}}(\beta_{B''}) X_{\ketbra{-\alpha_0}{\alpha_0}{}}(-\beta_{B''})\nonumber\\
&\;+\frac{1}{\pi}\int d^2\beta_D X_{\ketbra{-}{+}{}}(\beta_D) X_{\ketbra{+}{-}{}}(-\beta_D)\nonumber\\
& \left.\qquad\times \frac{1}{\pi}\int d^2\beta_{B''} X_{\ketbra{-\alpha_0}{\alpha_0}{}}(\beta_{B''}) X_{\ketbra{\alpha_0}{-\alpha_0}{}}(-\beta_{B''})\right].
\label{fidelity1_large_alpha}
\end{align}
All the integrals above are equal to $\pi$, giving $F_{self}^{(large)}=~1$. Similarly, for the case of small $\alpha_0$, $\bigchi_h^{(small)}(\beta_D,\beta_{B''})$ is given by Eq.~(\ref{hybrid_chi}). The DV mode in the $\{\ket{0},\ket{1}\}$ basis also exhibits a symmetry similar to that in Eqs.~(\ref{symmetrya}) and (\ref{symmetryb}). From such a symmetry, the fidelity of the state to itself can be simplified from sixteen terms to only four non-zero terms
\begin{align}
&F_{self}^{(small)} = \nonumber\\
&\frac{1}{4}\left[ \frac{1}{\pi}\int d^2\beta_D|X_{\ketbra{0}{0}{}}(\beta_D)|^2 \right.\nonumber\\
&\qquad\times \frac{1}{\pi}\int d^2\beta_{B''}|X_{\ketbra{1}{1}{}}(\tilde{\beta}_{B''})|^2 \nonumber\\
&\;+\frac{1}{\pi}\int d^2\beta_D|X_{\ketbra{1}{1}{}}(\beta_D)|^2 \nonumber\\
&\qquad\times \frac{1}{\pi}\int d^2\beta_{B''}|X_{\ketbra{0}{0}{}}(\tilde{\beta}_{B''})|^2\nonumber\\
&\;+\frac{1}{\pi}\int d^2\beta_D X_{\ketbra{0}{1}{}}(\beta_D)X_{\ketbra{1}{0}{}}(-\beta_D) \nonumber\\
&\qquad\times \frac{1}{\pi}\int d^2\beta_{B''} X_{\ketbra{1}{0}{}}(\tilde{\beta}_{B''}) X_{\ketbra{0}{1}{}}(-\tilde{\beta}_{B''})\nonumber\\
&\;+\frac{1}{\pi}\int d^2\beta_D X_{\ketbra{1}{0}{}}(\beta_D) X_{\ketbra{0}{1}{}}(-\beta_D)\nonumber\\
& \left.\qquad\times \frac{1}{\pi}\int d^2\beta_{B''} X_{\ketbra{0}{1}{}}(\tilde{\beta}_{B''}) X_{\ketbra{1}{0}{}}(-\tilde{\beta}_{B''})\right].
\label{fidelity1_small_alpha}
\end{align}
All the integrals above are equal to $\pi$, again giving $F_{self}^{(small)}=~1$.

When the CV mode of the hybrid entangled state is teleported, mode $C$ will be teleported to mode $B''$. Assuming a teleportation gain of $g=1$,\footnote{$g=1$ was found to be the optimal gain for symmetric channels with regard to the logarithmic negativity \cite{takeda2013gaintuning}\cite{do2019hybrid}.} the teleportation output is given by \cite{farias2009thesis}\cite{dellanno2018nongaussian}
\beq
\bigchi_{tel}(\beta_D, \beta_{B''}) = \bigchi_h(\beta_D, \beta_{B''})\bigchi_{TMSV}^{(T_A,T_B)}(\beta_{B''}^*,\beta_{B''}),
\label{teleportation_characteristic}
\eeq
where the channel is given by the attenuated TMSV state\footnote{More generally, see  \cite{marian2006continuous} for how teleportation can operate under any type of CV teleportation channel.} $\bigchi_{TMSV}^{(T_A,T_B)}(\beta_{B''}^*,\beta_{B''})$ in Eq.~(\ref{TMSV_tatb}).
The fidelity of a hybrid entangled state after teleportation can be calculated by applying Eq.~(\ref{eq_fidelity}) \cite{farias2009thesis,seshadreesan2015nongaussian,dellanno2018nongaussian}
\begin{align}
F_{tel}
&=\frac{1}{\pi}\iint d^2\beta_D d^2\beta_{B''}\bigchi_h(\beta_D,\beta_{B''})\bigchi_h(-\beta_{D},-\beta_{B''})\nonumber\\
&\qquad \qquad\times \bigchi_{TMSV}^{(T_A,T_B)}(-\beta_{B''}^*,-\beta_{B''}),
\end{align}
where, by applying Eqs.~(\ref{vac_chi}), (\ref{TMSV_chi}), and (\ref{TMSV_tatb}), we have
\beq
\bigchi_{TMSV}^{(T_A,T_B)}(-\beta_{B''}^*,-\beta_{B''}) = \exp \left[-\sigma|\beta_{B''}|^2\right],
\label{TMSV_minus}
\eeq
with $\sigma$ defined in Eq.~(\ref{sigma_general}). When there is no loss ($T_A=T_B=1$), we have $\sigma = e^{-2r}$ \cite{seshadreesan2015nongaussian}, where $r$ is again the initial squeezing. When the two channels are symmetric ($T_A=T_B=T$), we have \beq
\sigma =T e^{-2r} + (1-T).
\label{sigma_T}
\eeq
In order to calculate the fidelity, we perform a summation of integrations similar to that in Eqs.~(\ref{fidelity1_large_alpha}) and (\ref{fidelity1_small_alpha}).
In the following, we will calculate the fidelity for the cases when $\alpha_0$ is large or small, respectively.
For hybrid states with intermediate sizes, the detailed calculations can be found in appendix~\ref{appendix_cat_exact}.

\paragraph{Large cat state}
\label{large_fidelity}
When calculating the fidelity of the large hybrid state after teleportation, similar to Eqs.~(\ref{symmetrya}), (\ref{symmetryb}), and (\ref{fidelity1_large_alpha}), we obtain sixteen terms, which can be simplified to four non-zero terms by applying the DV-mode symmetry, giving
\begin{align}
&F_{tel}^{(large)} = \nonumber\\
&\frac{1}{4}\left[\vphantom{\frac{1}{4}}\right. \frac{1}{\pi}\int d^2\beta_D|X_{\ketbra{+}{+}{}}(\beta_D)|^2 \nonumber\\
&\times \frac{1}{\pi}\int d^2\beta_{B''}|X_{\ketbra{\alpha_0}{\alpha_0}{}}(\beta_{B''})|^2 \bigchi_{TMSV}^{(T_A,T_B)}(-\beta_{B''}^*,-\beta_{B''}) \nonumber\\
&+\frac{1}{\pi}\int d^2\beta_D|X_{\ketbra{-}{-}{}}(\beta_D)|^2 \nonumber\\
&\times \frac{1}{\pi}\int d^2\beta_{B''}|X_{\ketbra{-\alpha_0}{-\alpha_0}{}}(\beta_{B''})|^2  \bigchi_{TMSV}^{(T_A,T_B)}(-\beta_{B''}^*,-\beta_{B''}) \nonumber\\
&+\frac{1}{\pi}\int d^2\beta_D X_{\ketbra{+}{-}{}}(\beta_D)X_{\ketbra{-}{+}{}}(-\beta_D) \nonumber\\
&\times \frac{1}{\pi}\int d^2\beta_{B''} X_{\ketbra{\alpha_0}{-\alpha_0}{}}(\beta_{B''}) X_{\ketbra{-\alpha_0}{\alpha_0}{}}(-\beta_{B''})  \nonumber\\
&\qquad\quad \times\bigchi_{TMSV}^{(T_A,T_B)}(-\beta_{B''}^*,-\beta_{B''})\nonumber\\
&+{\frac{1}{\pi}}\int d^2\beta_D X_{\ketbra{-}{+}{}}(\beta_D) X_{\ketbra{+}{-}{}}(-\beta_D)\nonumber\\
& \times \frac{1}{\pi}\int d^2\beta_{B''} X_{\ketbra{-\alpha_0}{\alpha_0}{}}(\beta_{B''}) X_{\ketbra{\alpha_0}{-\alpha_0}{}}(-\beta_{B''})\nonumber\\
 & \qquad\quad \times\left.\bigchi_{TMSV}^{(T_A,T_B)}(-\beta_{B''}^*,-\beta_{B''})\right].
\end{align}
The integrations over mode $D$ are still equal to $\pi$. For the integrations  over mode $B''$, after applying Eq.~(\ref{TMSV_minus}), we define the following four terms and calculate them using Mathematica
\begin{align}
f_{\pm\pm\alpha_0} &= \frac{1}{\pi}\int d^2\beta_{B''}|X_{\ketbra{\pm\alpha_0}{\pm\alpha_0}{}}(\beta_{B''})|^2 e^{-\sigma|\beta_{B''}|^2} \nonumber\\
&= \frac{1}{\pi}\int d^2\beta_{B''}e^{-|\beta|^2} e^{-\sigma|\beta_{B''}|^2}= \frac{1}{1+\sigma},\\
f_{\pm\mp\alpha_0} &=\frac{1}{\pi}\int d^2\beta_{B''} X_{\ketbra{\pm\alpha_0}{\mp\alpha_0}{}}(\beta_{B''}) X_{\ketbra{\mp\alpha_0}{\pm\alpha_0}{}}(-\beta_{B''})\nonumber\\
&\qquad \times e^{-\sigma|\beta_{B''}|^2}\nonumber\\
&= \frac{1}{\pi}\int d^2\beta_{B''} e^{-|\beta_{B''}\pm 2\alpha_0|^2}e^{-\sigma|\beta_{B''}|^2} \nonumber\\
&=\frac{1}{1+\sigma} \exp\left(-\frac{4\alpha_0^2\sigma}{1+\sigma}\right).
\end{align}
So the fidelity after teleportation is
\begin{align}
F_{tel}^{(large)} &= \frac{1}{4}(f_{++\alpha_0}+ f_{--\alpha_0}+ f_{+-\alpha_0}+f_{-+\alpha_0}) \nonumber\\
&= \frac{1}{2(1+\sigma)}\left[1+\exp\left(-\frac{4\alpha_0^2\sigma}{1+\sigma}\right)\right].
\label{fidelity_large_alpha}
\end{align}

\paragraph{Small cat state}

\begin{figure*}[h!]
    \centering

		\begin{subfigure}[b]{0.46\linewidth}
		        \includegraphics[width=\linewidth]{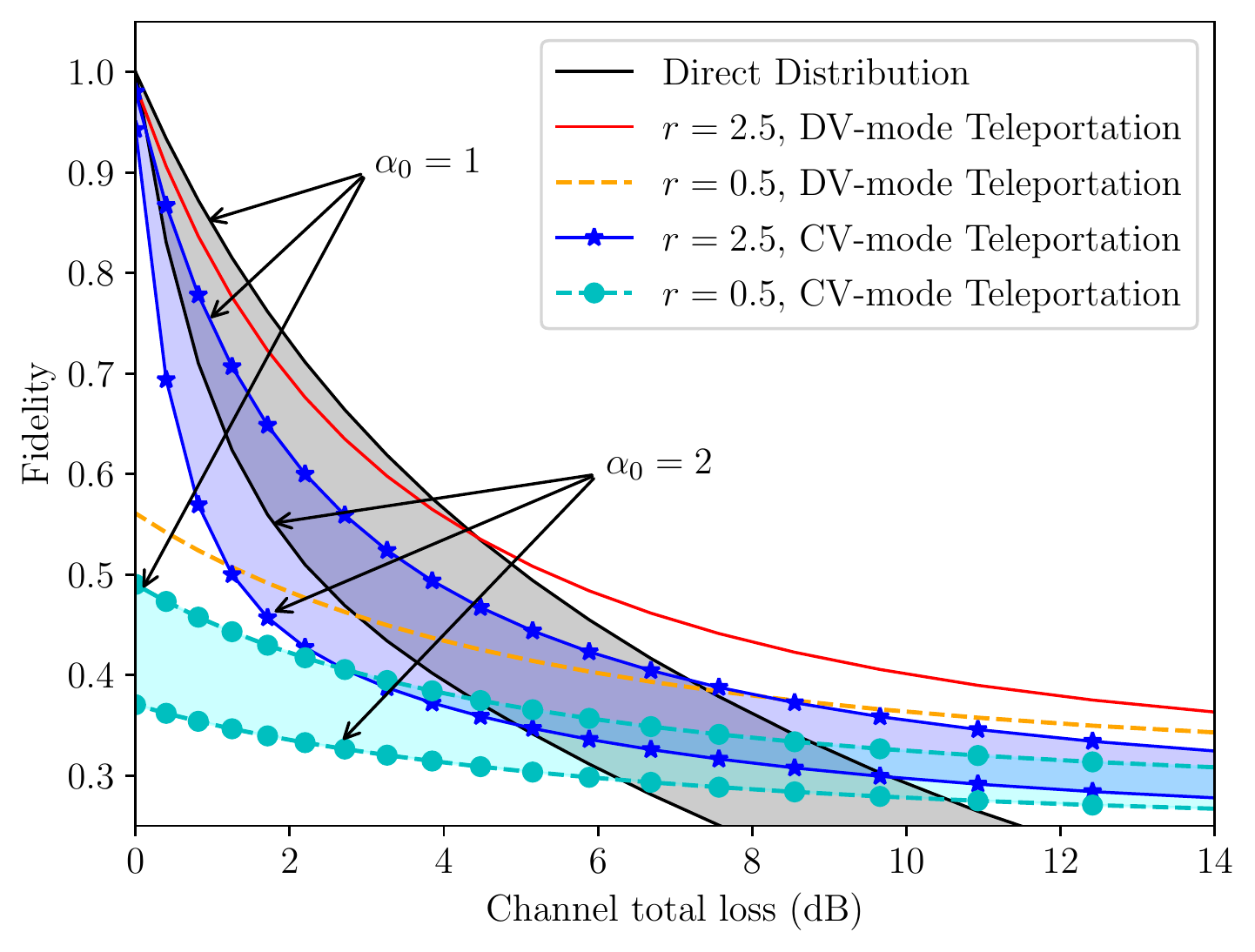}
						\caption{$\alpha_0 = 1 \rightarrow 2$}
				\label{fidelity_large}
		\end{subfigure}
				\begin{subfigure}[b]{0.46\linewidth}
        \includegraphics[width=\linewidth]{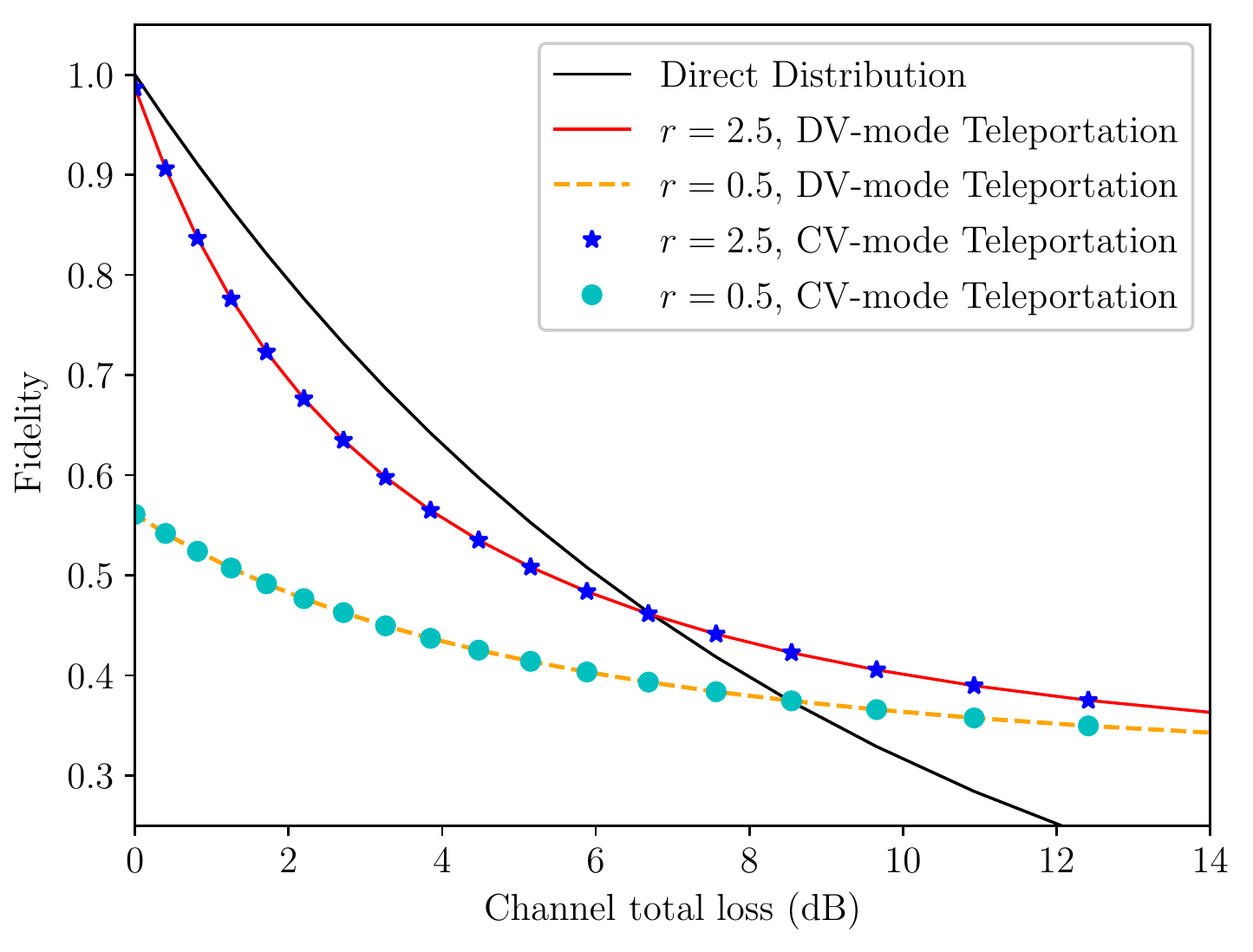}
				\caption{$\alpha_0 < 0.5$}
				\label{fidelity_small}
		\end{subfigure}
		\caption{The fidelity is plotted for direct distribution (black), or teleportation of the the DV/CV mode (red-orange/blue-navy). The channels are assumed to be symmetric ($T_A = T_B$), while  the initial two-mode squeezing parameter $r$ is set to either 2.5 or 0.5, and the teleportation gain is set to $g=1$. In general, the fidelity of DV-mode teleportation stays the same regardless of $\alpha_0$. (a) For $\alpha_0 = 1 \rightarrow 2$, the fidelities of direct distribution and CV-mode teleportation vary in the shaded ranges, with the upper bound corresponding to $\alpha_0=1$, and the lower bound corresponding to $\alpha_0=2$. (b) For $\alpha_0< 0.5$, the fidelities of the three schemes stay approximately unchanged, providing an upper bound for the fidelity with respect to $\alpha_0$.}
		\label{fidelity}
\end{figure*}

When calculating the fidelity of the small hybrid state after CV-mode teleportation, we arrive at sixteen different terms. Similar to Eq.~(\ref{fidelity1_small_alpha}), by applying the DV-mode symmetry, twelve of these terms become zero, leaving the following four terms
\begin{align}
&F_{tel}^{(small)} = \nonumber\\
&\frac{1}{4}\left[ \frac{1}{\pi}\int d^2\beta_D|X_{\ketbra{0}{0}{}}(\beta_D)|^2 \right.\nonumber\\
&\qquad\times \frac{1}{\pi}\int d^2\beta_{B''}|X_{\ketbra{1}{1}{}}(\tilde{\beta}_{B''})|^2 \bigchi_{TMSV}^{(T_A,T_B)}(-\beta_{B''}^*,-\beta_{B''})  \nonumber\\
&\;+\frac{1}{\pi}\int d^2\beta_D|X_{\ketbra{1}{1}{}}(\beta_D)|^2 \nonumber\\
&\qquad\times \frac{1}{\pi}\int d^2\beta_{B''}|X_{\ketbra{0}{0}{}}(\tilde{\beta}_{B''})|^2\bigchi_{TMSV}^{(T_A,T_B)}(-\beta_{B''}^*,-\beta_{B''}) \nonumber\\
&\;+\frac{1}{\pi}\int d^2\beta_D X_{\ketbra{0}{1}{}}(\beta_D)X_{\ketbra{1}{0}{}}(-\beta_D) \nonumber\\
&\qquad\times \frac{1}{\pi}\int d^2\beta_{B''} X_{\ketbra{1}{0}{}}(\tilde{\beta}_{B''}) X_{\ketbra{0}{1}{}}(-\tilde{\beta}_{B''})\nonumber\\
&\qquad\qquad\quad \times\bigchi_{TMSV}^{(T_A,T_B)}(-\beta_{B''}^*,-\beta_{B''})\nonumber\\
&\;+\frac{1}{\pi}\int d^2\beta_D X_{\ketbra{1}{0}{}}(\beta_D) X_{\ketbra{0}{1}{}}(-\beta_D)\nonumber\\
& \qquad\times \frac{1}{\pi}\int d^2\beta_{B''} X_{\ketbra{0}{1}{}}(\tilde{\beta}_{B''}) X_{\ketbra{1}{0}{}}(-\tilde{\beta}_{B''})\nonumber\\
&\qquad\qquad\quad \times\left.\bigchi_{TMSV}^{(T_A,T_B)}(-\beta_{B''}^*,-\beta_{B''})\right].
\label{fidelity_small_alpha}
\end{align}
The integrations over mode $D$ still give $\pi$, for the integrations over mode $B''$, we apply Eq.~(\ref{TMSV_minus}) and define
\begin{align}
f_{00}&=\frac{1}{\pi}\int d^2\beta_{B''}|X_{\ketbra{0}{0}{}}(\tilde{\beta}_{B''})|^2 e^{-\sigma|\beta_{B''}|^2} = \frac{1}{\sqrt{\tau}},\nonumber\\
f_{11}&=\frac{1}{\pi}\int d^2\beta_{B''}|X_{\ketbra{1}{1}{}}(\tilde{\beta}_{B''})|^2 e^{-\sigma|\beta_{B''}|^2} \nonumber\\
&= \frac{2+\sigma^2+2\sigma^4+4\sigma(1+\sigma^2)\cosh(2s)+3\sigma^2\cosh(4s)}{2\sqrt{e^{-2s}+\sigma}(e^{2s}+\sigma)^{5/2}(e^{-2s}+\sigma)^2},
\nonumber\\
f_{10}&=\frac{1}{\pi}\int d^2\beta_{B''}X_{\ketbra{1}{0}{}}(\tilde{\beta}_{B''})X_{\ketbra{0}{1}{}}(\tilde{\beta}_{B''}) e^{-\sigma|\beta_{B''}|^2}  \nonumber\\
&= \frac{1+\sigma\cosh(2s)}{\tau^{3/2}},\nonumber\\
f_{01}&=\frac{1}{\pi}\int d^2\beta_{B''}X_{\ketbra{0}{1}{}}(\tilde{\beta}_{B''})X_{\ketbra{1}{0}{}}(-\tilde{\beta}_{B''}) e^{-\sigma|\beta_{B''}|^2} \nonumber\\
&=f_{10},
\end{align}
where $\tau = 1 + \sigma^2 +2\sigma\cosh(2s)$, with $s$ being the single-mode squeezing parameter in Eq.~(\ref{s}).
The final fidelity of hybrid entanglement after teleportation is given by
\beq
F_{tel}^{(small)} = \frac{1}{4}\left(f_{00}+f_{11}+ f_{10} + f_{01}\right),
\label{fidelity_cv}
\eeq
which tends to unity as the channel loss approaches zero ($T_A=T_B\rightarrow 1$).
The detailed simulation results are presented in the next section.


\section{Results}
\label{result}

In this section, we aim to compare the long-distance distribution of hybrid entanglement using two different schemes: one scheme (in Fig.~\ref{direct_setup}) uses the direct distribution from the satellite, with one terrestrial node receiving the DV mode and the other node the CV mode; and one scheme (in Fig.~\ref{teleported_setup}) uses the TMSV channel created by the satellite to teleport one mode of an initially localised hybrid state from one terrestrial node to another. We compare the two schemes in terms of the fidelity and logarithmic negativity.

We simulate the hybrid entangled state in Eq.~(\ref{hybrid_CD}) with different real values of $\alpha_0$,\footnote{
The same simulation was run with complex values of $\alpha_0$ as well. However, for both direct distribution and DV-mode teleportation, we found that the fidelity is the highest when the initial cat states are small and real.
} especially in the cases of small cat states ($\alpha_0~<~0.5$) and large cat states ($\alpha_0~>~1$).
For DV-mode teleportation, when calculating the summation in Eq.~(\ref{teleported_output}), we stop at a value of $k=k_{max}$. Since the averaged photon number is given by $\alpha_0^2$, we see that $k_{max}$ should be at least $\alpha_0^2$ \cite{parker2018thesis}. Practically, to find $k_{max}$, we start with a high value, then slowly reduce $k_{max}$ while checking that $\tr\left(\ketbra{cat_\pm}{cat_\pm}{}\right) = 1 \pm \delta$, where $\delta$ is the tolerance. In the following simulations, we use the values of $\delta=10^{-14}$ and $k_{max} \approx 30$.

In our first simulation, we assume that the two down-link channels are symmetric, i.e., $T_A = T_B = T$. The total loss of two channels can be calculated by $-10 \log_{10}(T^2)$ (dB). The teleportation gain is set to $g=1$, which is the optimal gain when $T_A=T_B$ \cite{takeda2013gaintuning}\cite{do2019hybrid}.
The fidelity is calculated by Eqs.~(\ref{arbitrary_alpha}), (\ref{fidelity_rho}), (\ref{teleported_output}), (\ref{fidelity_large_alpha}), and (\ref{fidelity_cv}).  Fig.~\ref{fidelity_large} shows the case of large cat states where we selectively adopted values of  $\alpha_0$ in the range $1-2$ ($\alpha_0 = 1\rightarrow 2$).
We first vary the initial squeezing $r$ of the TMSV channel and find that the fidelity of teleportation increases as $r$ increases, then saturates to its maximum value when $r\approx 2.5$, which we will refer to as the optimal squeezing parameter. Thus, in this figure, we plot the fidelity for $r=2.5$ and a lower value of $r=0.5$. We found that, for higher loss, the effect of the initial squeezing $r$ on teleportation results becomes less significant. This implies that for long-distance communications, we need not spend too much experimental resources on enhancing the channel squeezing.

We can also see from Fig.~\ref{fidelity_large} that the fidelity of DV-mode teleportation stays the same  regardless of $\alpha_0$, while the fidelities of CV-mode teleportation and direct distribution decrease when $\alpha_0$ increases. In particular, for $r=2.5$, our simulation shows that when $\alpha_0>1.2$, the fidelity of DV-mode teleportation is always higher than direct distribution, and the difference further increases with increasing $\alpha_0$ and higher channel loss.
When $\alpha_0$ decreases below 1.2, the fidelity of direct distribution starts to become higher than DV-mode teleportation at low channel loss. However, as the loss increases, the fidelity of direct distribution quickly decreases, while the fidelity of DV-mode teleportation decreases slower then saturates to a higher value. The loss level where the crossover occurs (which we will henceforth refer to as the `crossover loss') increases as $\alpha_0$ decreases, and varies from 0dB go 5dB when $\alpha_0$ varies from 1.2 to 1.
This crossover can be explained by how the channel loss enters the calculations of the direct distribution scheme (via both DV and CV loss components) as compared to the teleportation channel (via two CV loss components). Ultimately, our detailed mathematical models offer the explanation of these findings.

\begin{figure*}[h!]
    \centering
		\begin{subfigure}[b]{0.32\linewidth}
        \includegraphics[width=\linewidth]{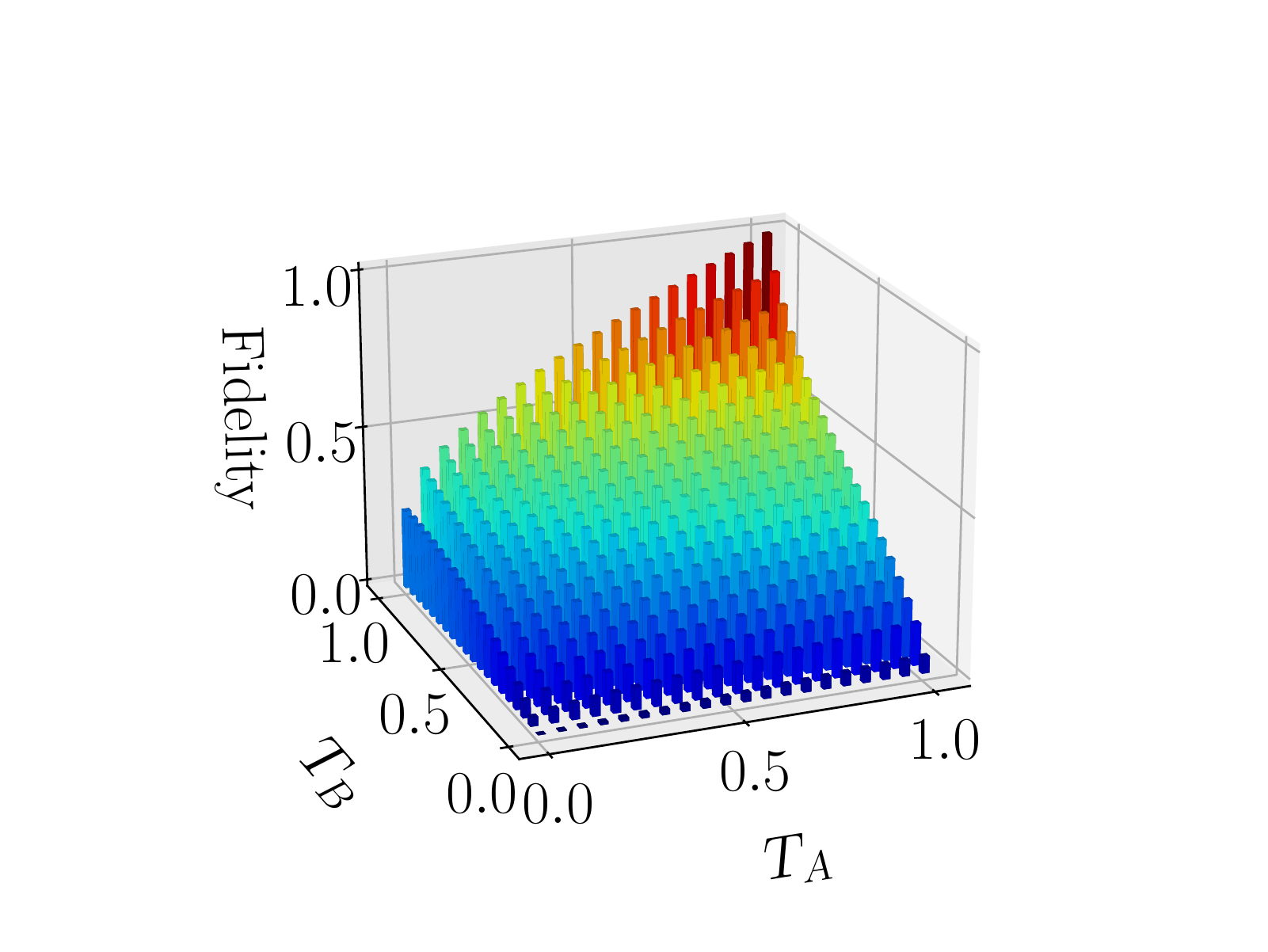}
        \caption{Direct Distribution}
        \label{3d_f_direct}
    \end{subfigure}
    \begin{subfigure}[b]{0.32\linewidth}
        \includegraphics[width=\linewidth]{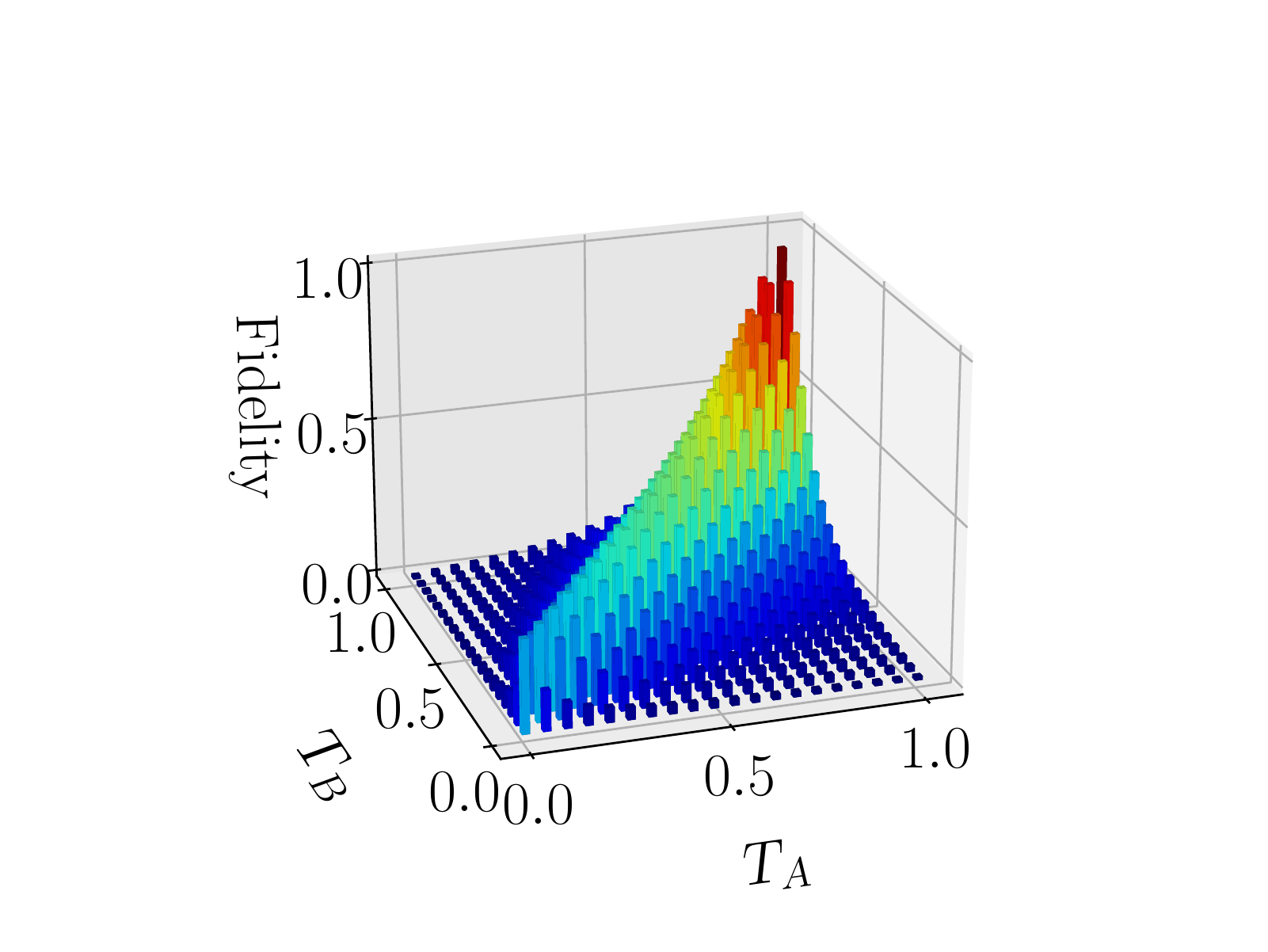}
        \caption{DV-mode Teleportation, $g=1$}
        \label{3d_f_teleport}
    \end{subfigure}
    \begin{subfigure}[b]{0.32\linewidth}
        \includegraphics[width=\linewidth]{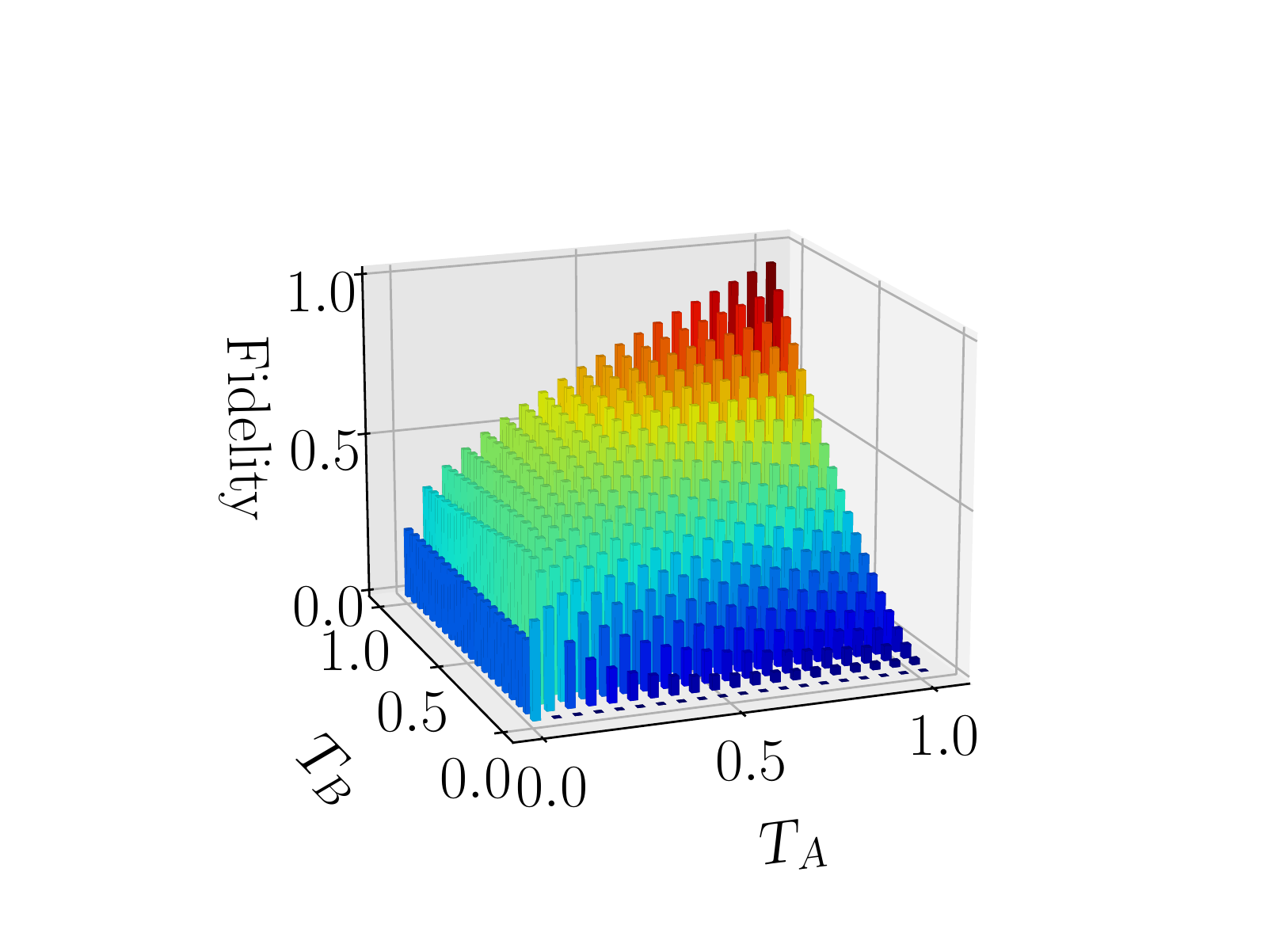}
        \caption{DV-mode Teleportation, $g=\sqrt{\frac{T_B}{T_A}}$}
        \label{3d_f_teleport_g}
    \end{subfigure}
\caption{The fidelities of direct distribution (a) and DV-mode teleportation (b,c) is plotted where $T_A$ and $T_B$ are independently varied. We fixed $\alpha_0=1.5$, $r=2.5$, and $g=1$ (b) or $g=\sqrt{T_B/T_A}$ (c).
		}		
\label{3D}
\end{figure*}
\begin{figure*}[h!]
\centering
		\begin{subfigure}[b]{0.32\linewidth}
        \includegraphics[width=\linewidth]{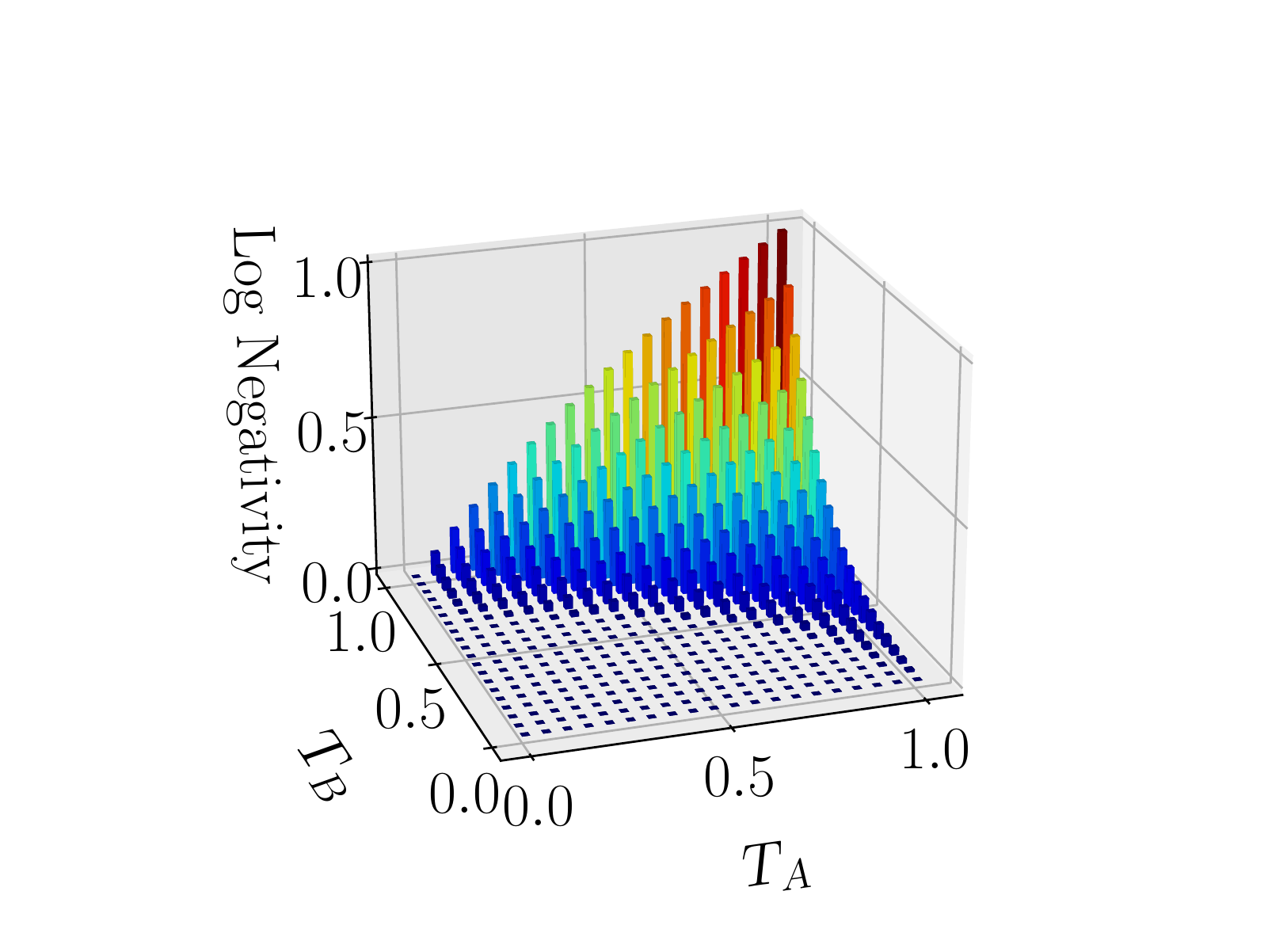}
        \caption{Direct Distribution}
        \label{3d_ln_direct}
    \end{subfigure}
    \begin{subfigure}[b]{0.32\linewidth}
        \includegraphics[width=\linewidth]{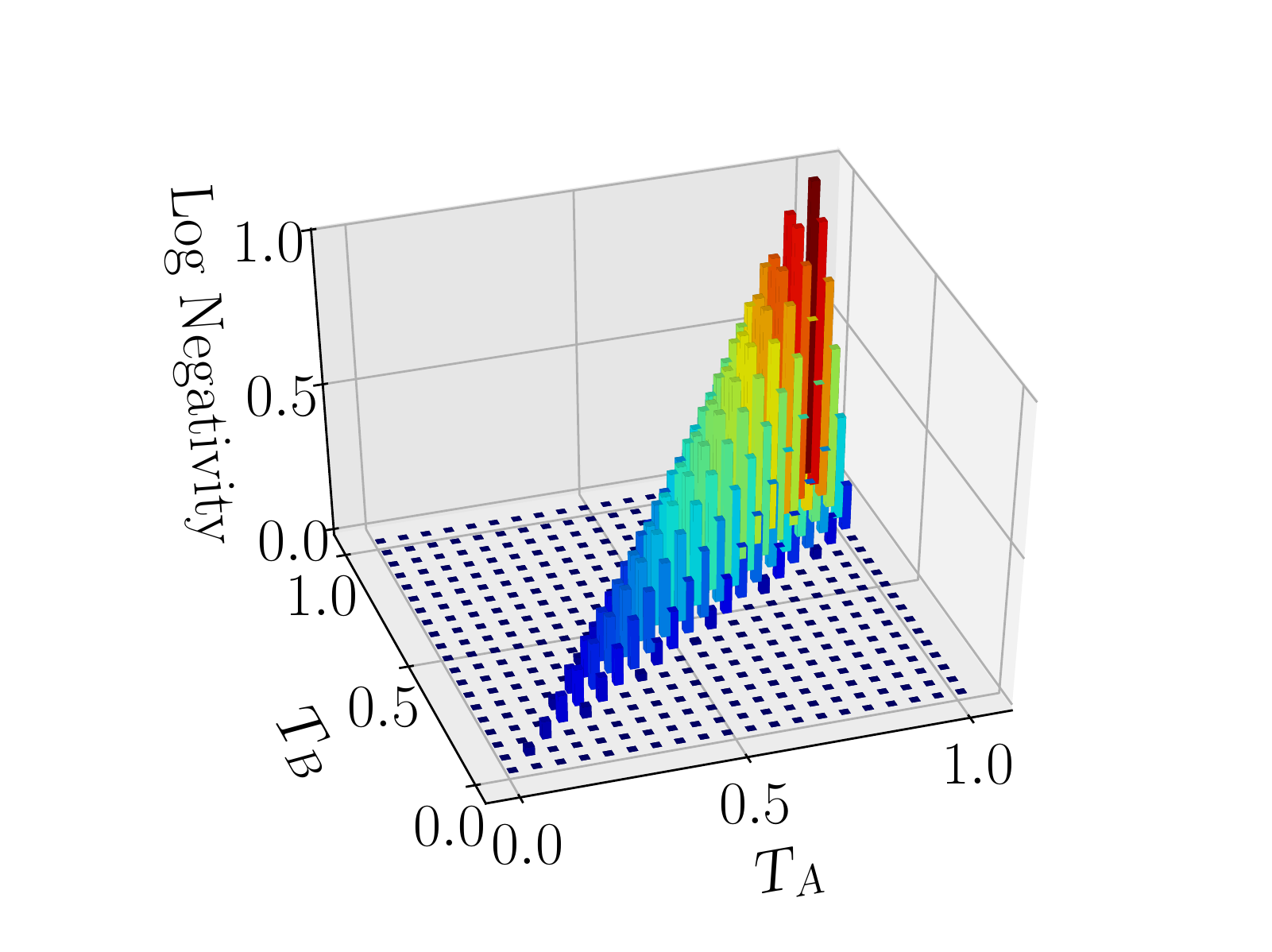}
        \caption{DV-mode Teleportation, $g=1$}
        \label{3d_ln_teleport}
\end{subfigure}
    \begin{subfigure}[b]{0.32\linewidth}
        \includegraphics[width=\linewidth]{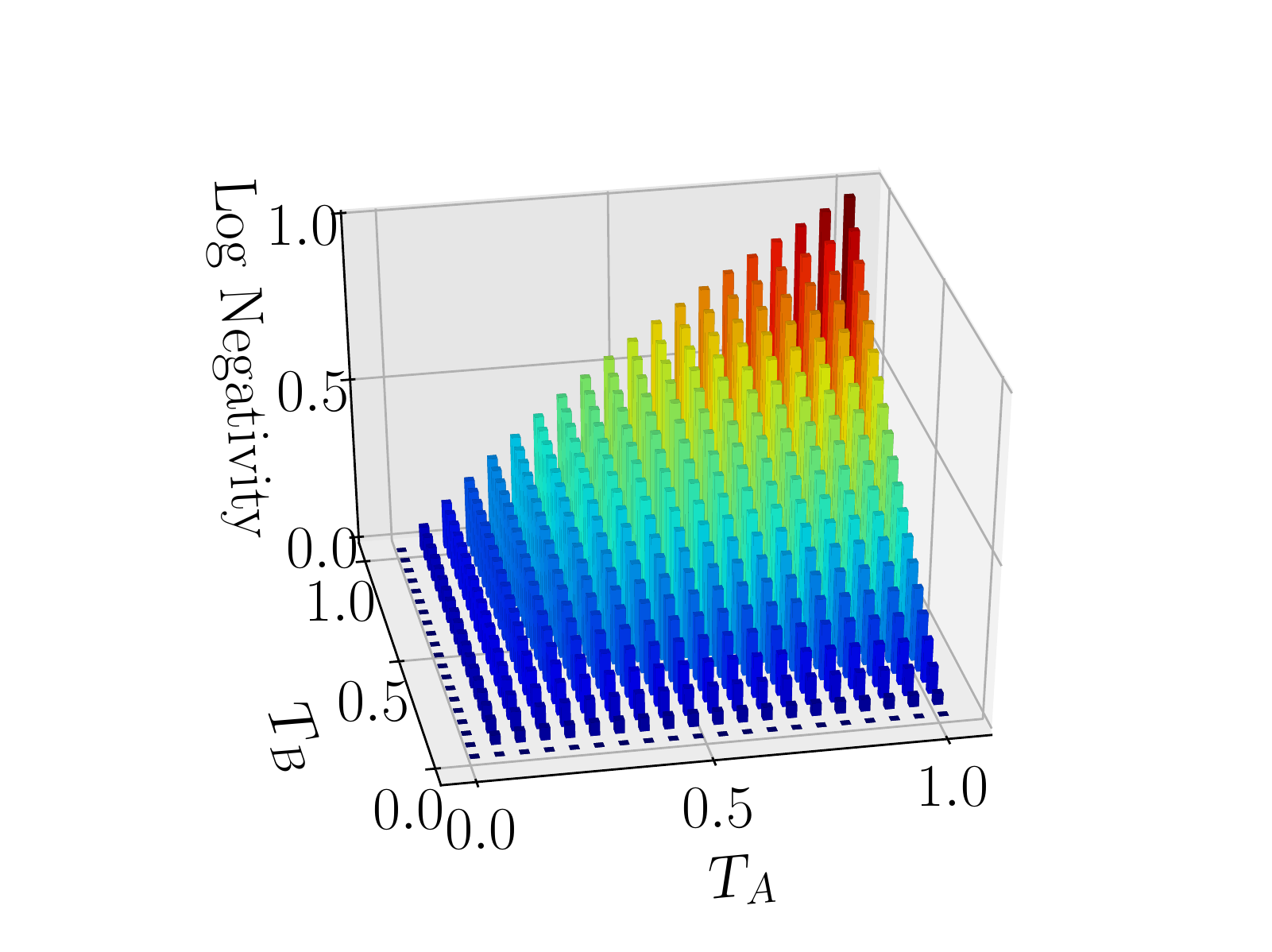}
        \caption{DV-mode Teleportation, $g=\sqrt{\frac{T_B}{T_A}}$}
        \label{3d_ln_teleport_g}
    \end{subfigure}
    \caption{The logarithmic negativity of direct distribution (a) and DV-mode teleportation (b,c) with $\alpha_0=1.5$, $r=2.5$, and $g=1$ (b) or $\sqrt{\frac{T_B}{T_A}}$ (c).
		}
\label{3D_LN}
\end{figure*}

In addition, in Fig.~\ref{fidelity_large}, under all loss conditions, teleporting the DV mode of the hybrid entangled state is always better or equal to teleporting the CV mode. The difference becomes more significant when $\alpha_0$ increases. Our results can be explained by the notion that for a given squeezing level and a given loss level in the CV teleportation channel, the teleportation of larger Fock states will provide for less fidelity than smaller Fock states. Only in the limit of zero loss would this not be the case. As such, superpositions of Fock states weighted to excitations of small photon numbers should be `easier' to teleport. Our DV state can be considered loosely as such a superposition of Fock states.

Fig.~\ref{fidelity_small} shows the case of small cat states where we selectively simulated $\alpha_0 = 0.1 \rightarrow 0.5$. In this case, we can see that the fidelities of the three schemes all remain almost unchanged, and the difference cannot be seen from the curves in the plot. For losses below 7dB, we can also see that direct distribution gives a higher fidelity than teleportation. However, when the loss increases above 7dB, teleportation is always better. In addition, for all loss conditions, we find that it does not matter to teleport either the DV or CV mode of hybrid entanglement, since the output fidelity is the same. This symmetry can be explained by the fact that when $\alpha_0<0.5$, Eq.~(\ref{s}) gives $s<0.1$, so that the SMSV and 1PS states in Eq.~(\ref{cat_approximation}) are very lightly squeezed. Intuitively, we can understand that when the cat states become small, their average photon numbers approach either 0 or 1. As a result, the cat states approach $\ket{0}$ and $\ket{1}$, and the approximated hybrid entangled state in Eq.~(\ref{approx_hybrid}) approaches a DV entangled state, which is spatially symmetric.  In general, we can see that lower photon-number states, especially qubit states, are less sensitive to channel loss. Thus, the case of small $\alpha_0$ provides the upper bounds for the fidelities of the three schemes.

For cat states of intermediate sizes where $0.5<\alpha_0<1$, our simulations show that the fidelity of DV-mode teleportation still stays unchanged, while the fidelities of direct distribution and CV-mode teleportation decrease monotonically as $\alpha_0$ increases. For this intermediate range, the fidelities  naturally stay in between the bounds set by large and small $\alpha_0$.

To study asymmetric channels, we next vary $T_A$ and $T_B$ independently, while plotting the 3D map for the fidelity. Since large cat states are more useful for quantum computations, in this simulation, we fix $\alpha_0=1.5$ and $r=2.5$, respectively. Since our simulations have shown that DV-mode is always preferable over CV-mode teleportation, especially for large $\alpha_0$, we will only plot the fidelity for direct distribution and DV-mode teleportation.
From the plot of direct distribution (Fig.~\ref{3d_f_direct}), we see that the fidelity decreases monotonically with $T_A$ and $T_B$. Especially, the fidelity decreases faster with the loss in the CV mode (mode $B$). For DV-mode teleportation, in order to study the effect of gain-tuning, we set the gain to either $g=1$ or $g=\sqrt{T_B/T_A}$.\footnote{This $g$ value was previously found to be optimal with
regard to logarithmic negativity (see section~\ref{section_dv_teleportation} and \cite{do2019hybrid}).} As can be seen in Fig.~\ref{3d_f_teleport}, when $g=1$, the output attains the highest level of entanglement when the channels are symmetric $T_A=T_B$, but quickly drops to near zero when $T_A\neq T_B$. In Fig.~\ref{3d_f_teleport_g}, we can see that tuning the teleportation gain to $g=\sqrt{T_B/T_A}$ significantly improves the fidelity for asymmetric channels \cite{do2019hybrid}.
During gain tuning, it is especially important to maintain a high transmissivity in Bob's down-link channel, since when $T_B$ is high, the fidelity remains reasonably high for all different values of $T_A$.

In the next simulation, we use the same parameters as the previous one, but we plot the logarithmic negativity using Eq.~(\ref{ELN_rho}). From the results in Fig.~\ref{3D_LN}, we see that the logarithmic negativity follows the same trend as the fidelity, however, the former decreases more sharply with low transmissivities and channel asymmetry as compared to the latter. We also see that tuning the gain to the optimal value of $g=\sqrt{T_B/T_A}$ significantly improves the quality of the teleportation outcomes.

We also perform a series of similar simulations where the initial hybrid entangled cat state in Eq.~(\ref{hybrid_ket}) is replaced by a hybrid coherent state of the form \cite{kreis2012classifying}\cite{sheng2013hybrid}
\begin{align}
 ( \ket{\alpha_0}_C\ket{0}_D + \ket{-\alpha_0}_C\ket{1}_D)/\sqrt{2}.
\end{align}
We find that for large $\alpha_0$, all our conclusions found for our cat hybrid state remain intact. Indeed, the detailed results show trends for both direct and teleported states very similar to those shown in  Fig.~\ref{fidelity_large}.
For small $\alpha_0$, clear differences in both the direct and teleported states can be identified relative to Fig.~\ref{fidelity_small}.
However, again we emphasize the large $\alpha_0$ states are the states of interest for almost all hybrid DV-CV quantum protocols.
States beyond large cat states remain worthy of study, particularly in regard to upper bounds on the teleported fidelities.




\section{Conclusion}
\label{conclusion}
Hybrid technologies play an important role in interfacing mixed quantum protocols, which will in turn help accelerate the global development of quantum networks.
In this work, we studied, for the first time, the teleportation of either mode of a hybrid entangled state via a lossy TMSV channel, and compared such teleportation to the case where the hybrid entanglement is generated on board a satellite and directly distributed to Earth. Due to the complexity of the hybrid entangled state, especially in the cross-diagonal matrices with multi-photon numbers, we resorted to  the characteristic function formalism. By identifying symmetries within this formalism, we then developed a novel mathematical framework to calculate the fidelity of the teleported hybrid  state.
Our results showed that when the initial cat states have $\alpha_0>1.2$, it was always preferable to teleport the DV mode of the hybrid entangled state.
For cat states with $\alpha_0<1.2$, direct distribution gave a higher fidelity than teleportation at low channel loss.
However, for losses higher than 7dB, the teleportation of the DV mode was always better, regardless of $\alpha_0$.
We note, losses above 7dB are expected for satellite-to-Earth channels for transceiver apertures of reasonable sizes.
For all loss conditions, we found CV-mode teleportation always gave a fidelity that is less than or equal to DV-mode teleportation, where the equality happens when the cat states become small ($\alpha_0<0.5$).
Similar results to the above were found for a hybrid state in which the cat states were replaced by large coherent states of opposite phases.
Our results will be important for next-generation heterogeneous quantum communication networks, whose teleportation resource is sustained by standard TMSV entanglement distribution from LEO satellites.

\bibliographystyle{mybst2.bst}
\bibliography{myBib_p}

\begin{appendices}

\section{CV-mode teleportation of an exact hybrid entangled state}
\label{appendix_cat_exact}
This appendix uses the exact mathematical form of the hybrid entangled state (Eq.~(\ref{hybrid_CD})) to calculate the characteristic function and fidelity after CV-mode teleportation. The mathematical model below is especially useful for hybrid states with intermediate sizes ($0.5<\alpha_0<1$), where a simple approximation does not exist.

The characteristic function of the exact hybrid entangled state is found by substituting Eq.~(\ref{hybrid_CD}) into Eqs.~(\ref{characteristic_function_define}) and (\ref{X_function}), giving
\begin{align}
\bigchi_{h}^{(exact)}(\beta_D,\beta_C)
&= \frac{1}{2}\left[X_{\ketbra{0}{0}{}}(\beta_D) X_{\ketbra{cat_-}{cat_-}{}}(\beta_C) \right.\nonumber\\
&\qquad + X_{\ketbra{1}{1}{}}( \beta_D) X_{\ketbra{cat_+}{cat_+}{}}(\beta_C) \nonumber\\
&\qquad+X_{\ketbra{0}{1}{}}(\beta_D) X_{\ketbra{cat_-}{cat_+}{}}(\beta_C) \nonumber\\
&\qquad + \left. X_{\ketbra{1}{0}{}}(\beta_D) X_{\ketbra{cat_+}{cat_-}{}}( \beta_C)\right].
\end{align}


To find the fidelity after teleporting the CV mode, we follow the calculations in section~\ref{section_fidelity_cv}, giving
\begin{align}
&F_{tel}^{(exact)} = \frac{1}{4}\left[ \frac{1}{\pi}\int d^2\beta_D|X_{\ketbra{0}{0}{}}(\beta_D)|^2 \right.\nonumber\\
&\times \frac{1}{\pi}\int d^2\beta_{B''}|X_{\ketbra{cat_-}{cat_-}{}}(\beta_{B''})|^2  \bigchi_{TMSV}^{(T_A,T_B)}(-\beta_{B''}^*,-\beta_{B''}) \nonumber\\
&+\frac{1}{\pi}\int d^2\beta_D|X_{\ketbra{1}{1}{}}(\beta_D)|^2 \nonumber\\
&\times \frac{1}{\pi}\int d^2\beta_{B''}|X_{\ketbra{cat_+}{cat_+}{}}(\beta_{B''})|^2   \bigchi_{TMSV}^{(T_A,T_B)}(-\beta_{B''}^*,-\beta_{B''}) \nonumber\\
&+\frac{1}{\pi}\int d^2\beta_D X_{\ketbra{1}{0}{}}(\beta_D)X_{\ketbra{0}{1}{}}(-\beta_D) \nonumber\\
&\times \frac{1}{\pi}\int d^2\beta_{B''} X_{\ketbra{cat_+}{cat_-}{}}(\beta_{B''}) X_{\ketbra{cat_-}{cat_+}{}}(-\beta_{B''})  \nonumber\\
&\qquad\quad \times\bigchi_{TMSV}^{(T_A,T_B)}(-\beta_{B''}^*,-\beta_{B''})\nonumber\\
&+{\frac{1}{\pi}}\int d^2\beta_D X_{\ketbra{0}{1}{}}(\beta_D) X_{\ketbra{1}{0}{}}(-\beta_D)\nonumber\\
& \times \frac{1}{\pi}\int d^2\beta_{B''} X_{\ketbra{cat_-}{cat_+}{}}(\beta_{B''}) X_{\ketbra{cat_+}{cat_-}{}}(-\beta_{B''})\nonumber\\
 & \qquad\quad \times\left.\bigchi_{TMSV}^{(T_A,T_B)}(-\beta_{B''}^*,-\beta_{B''})\right].
\end{align}
The integrations over mode $D$ are still giving $\pi$, for the integrations in mode $B''$, after applying Eq.~(\ref{TMSV_minus}), the exact fidelity can be given by
\beq
F_{tel}^{(exact)} = \frac{1}{4}(f_{++cat} + f_{--cat} + f_{+-cat} + f_{-+cat}),
\label{f_exact}
\eeq
where
\begin{align}
f_{++cat} &= \frac{1}{\pi}\int d^2\beta_{B''} X_{\ketbra{cat_+}{cat_+}{}}(\beta_{B''})X_{\ketbra{cat_+}{cat_+}{}}(-\beta_{B''})\nonumber\\
&\qquad\quad\times e^{-\sigma|\beta_{B''}|^2},\nonumber\\
f_{--cat} &= \frac{1}{\pi}\int d^2\beta_{B''} X_{\ketbra{cat_-}{cat_-}{}}(\beta_{B''})X_{\ketbra{cat_-}{cat_-}{}}(-\beta_{B''})\nonumber\\
&\qquad\quad\times e^{-\sigma|\beta_{B''}|^2},\nonumber\\
f_{+-cat} &= \frac{1}{\pi}\int d^2\beta_{B''} X_{\ketbra{cat_+}{cat_-}{}}(\beta_{B''})X_{\ketbra{cat_-}{cat_+}{}}(-\beta_{B''})\nonumber\\
&\qquad\quad\times e^{-\sigma|\beta_{B''}|^2},\nonumber\\
f_{-+cat} &= \frac{1}{\pi}\int d^2\beta_{B''} X_{\ketbra{cat_-}{cat_+}{}}(\beta_{B''})X_{\ketbra{cat_+}{cat_-}{}}(-\beta_{B''})\nonumber\\
&\qquad\quad\times e^{-\sigma|\beta_{B''}|^2}.
\end{align}
In order to calculate the above integrations, we first apply Eq.~(\ref{cat}) to find
\begin{align}
&X_{\ketbra{cat_+}{cat_+}{}}(\beta)= \frac{1}{N_+^2}( \nonumber\\
&\quad X_{\ketbra{\alpha_0}{\alpha_0}{}}+X_{\ketbra{-\alpha_0}{-\alpha_0}{}}+X_{\ketbra{\alpha_0}{-\alpha_0}{}}+X_{\ketbra{-\alpha_0}{\alpha_0}{}}),\nonumber\\
&X_{\ketbra{cat_-}{cat_-}{}}(\beta) = \frac{1}{N_-^2}(\nonumber\\
&\quad X_{\ketbra{\alpha_0}{\alpha_0}{}}+X_{\ketbra{-\alpha_0}{-\alpha_0}{}}-X_{\ketbra{\alpha_0}{-\alpha_0}{}}-X_{\ketbra{-\alpha_0}{\alpha_0}{}}),\nonumber\\
&X_{\ketbra{cat_+}{cat_-}{}}(\beta) = \frac{1}{N_+N_-}( \nonumber\\
&\quad X_{\ketbra{\alpha_0}{\alpha_0}{}}-X_{\ketbra{-\alpha_0}{-\alpha_0}{}}-X_{\ketbra{\alpha_0}{-\alpha_0}{}}+X_{\ketbra{-\alpha_0}{\alpha_0}{}}),\nonumber\\
&X_{\ketbra{cat_-}{cat_+}{}}(\beta) = \frac{1}{N_+N_-}( \nonumber\\
&\quad X_{\ketbra{\alpha_0}{\alpha_0}{}}-X_{\ketbra{-\alpha_0}{-\alpha_0}{}}+X_{\ketbra{\alpha_0}{-\alpha_0}{}}-X_{\ketbra{-\alpha_0}{\alpha_0}{}}).
\label{X_function_cat}
\end{align}
We then define the following integrations
\begin{align}
f_{jlmn}&= \frac{1}{\pi}\int d^2\beta_{B''} X_{\ketbra{j\alpha_0}{l\alpha_0}{}}(\beta_{B''})X_{\ketbra{m\alpha_0}{n\alpha_0}{}}(-\beta_{B''})\nonumber\\
&\qquad\quad\times e^{-\sigma|\beta_{B''}|^2},
\end{align}
where $j,l,m,n\in\{+,-\}$. By substituting Eq.~(\ref{cf_coh}) into the above equation and using Mathematica to calculate, we find
\begin{align}
f_0 &= f_{\pm\pm\pm\pm} = \frac{1}{1+\sigma},\nonumber\\
f_1 &= f_{\pm\mp\mp\pm} = \frac{1}{1+\sigma} \exp\left(-\alpha_0 \frac{4\sigma}{1+\sigma}\right),\nonumber\\
f_2 &= f_{\pm\pm\mp\mp} = \frac{1}{1+\sigma} \exp\left(-\alpha_0 \frac{4}{1+\sigma}\right),\nonumber\\
f_3 &= f_{\pm\mp\pm\mp} = \frac{1}{1+\sigma} \exp\left(-4\alpha_0 \right),\nonumber\\
f_4 &= f_{\pm\pm\pm\mp}=f_{\mp\mp\pm\mp} \nonumber\\
&=f_{\pm\mp\pm\pm}=f_{\pm\mp\mp\mp}= \frac{1}{1+\sigma} \exp\left(-2\alpha_0 \right).
\label{f01234}
\end{align}
We can then find that
\begin{align}
f_{++cat} &= \frac{2}{N_+^4}(f_0+f_1+f_2+f_3 +4f_4),\nonumber\\
f_{--cat} &= \frac{2}{N_-^4}(f_0+f_1+f_2+f_3 -4f_4),\nonumber\\
f_{+-cat} &= f_{-+cat} =\frac{2}{N_+^2N_-^2}(f_0+f_1-f_2-f_3 ).
\label{f_pm_cat}
\end{align}
From Eqs.~(\ref{f_exact}), (\ref{f01234}), and (\ref{f_pm_cat}), we can find the exact fidelity of CV-mode teleportation. From our simulation, we can find that for cat states with intermediate sizes ($0.5<\alpha_0<1$), the fidelity stays between that of large and small cat states, and decreases monotonically as $\alpha_0$ increases.

\section{Direct distribution over a lossy channel}
\label{appendix_direct_distribution}
This section derives the attenuated hybrid entangled state after the direct distribution in Fig.~\ref{direct_setup} and section~\ref{direct_distribution}.
The original hybrid entangled state in Eq.~(\ref{hybrid_ket}), together with two auxiliary vacuum modes, can be written together as the four-mode state
\begin{align}
&\ket{\psi_{A\epsilon_A B \epsilon_B}} = \frac{1}{\sqrt{2}}\left[ \frac{\ket{\ao}_A  - \ket{-\ao}_A}{N_-}\ket{0}_{\epsilon_A}  \otimes \ket{0}_B \ket{0}_{\epsilon_B} \right.\nonumber\\
&+  \frac{\ket{\ao}_A + \ket{-\ao}_A}{N_+}\ket{0}_{\epsilon_A}\otimes \hat{a}^\dagger_B\ket{0}_B\ket{0}\epsilon_B\left.\vphantom{\frac{1}{2}}\right],
\end{align}
where $N_\pm$ is given in Eq.~(\ref{N_cat}).
The two lossy down-link channels are modeled by two beam splitters.
Let $t_l = \sqrt{T_l}$ and $r_l = \sqrt{1-T_l}$ with $l \in \{A,B\}$, after the two beam splitter transformations in Eqs.~(\ref{BStransformation_CV}) and (\ref{BStransformation_DV}), the state becomes
\begin{align}
&\ket{\psi_{A'\ea B' \eb}} = \frac{1}{\sqrt{2}}\left\{ \vphantom{\frac{1}{2}} \right. \nonumber\\
&\left[\frac{\ket{t_A \ao}_{A'} \ket{r_A \ao}_\ea - \ket{-t_A\ao}_{A'} \ket{-r_A\ao}_\ea}{N_-} \otimes \ket{0}_{B'} \ket{0}_\eb\right] \nonumber\\
&+ \left[\frac{\ket{t_A \ao}_{A'} \ket{r_A \ao}_\ea + \ket{-t_A\ao}_{A'} \ket{-r_A\ao}_\ea}{N_+}\right.\nonumber\\
& \quad\left.\left. \otimes \left(t_B\ket{1}_{B'} \ket{0}\eb + r_B\ket{0}_{B'}\ket{1}_\eb\right)
\vphantom{\frac{1}{2}}\right]\right\}.
\end{align}
The corresponding density matrix can be written as
\begin{align}
&\rho_{A'\ea B' \eb} = \frac{1}{2} \left\{\vphantom{\frac{r_B^2}{N_+^2}}\right.\nonumber\\
&\quad\;\ketbra{1}{1}{\eb}\otimes\left[\frac{r_B^2}{N_+^2} \ketbra{0}{0}{B'}\otimes(\rho_{++}+\rho_{--}+\rho_{-+}+\rho_{+-})\right]\nonumber\\
&+\ketbra{0}{0}{\eb}\otimes\left[\frac{t_B^2}{N_+^2} \ketbra{1}{1}{B'}\otimes(\rho_{++}+\rho_{--}+\rho_{-+}+\rho_{+-})\right.\nonumber\\
&\qquad \quad\;\;\,\quad +             \frac{1}{N_-^2}\;  \;   \ketbra{0}{0}{B'}\otimes(\rho_{++}+\rho_{--}-\rho_{-+}-\rho_{+-})\nonumber\\
&\qquad \quad \;\;\,+               \frac{t_B}{N_+ N_-} \ketbra{0}{1}{B'}\otimes(\rho_{++}-\rho_{--}+\rho_{-+}-\rho_{+-})\nonumber\\
&\qquad \quad \;\,+      \left.\frac{t_B}{N_+ N_-} \ketbra{1}{0}{B'}\otimes(\rho_{++}-\rho_{--}-\rho_{-+}+\rho_{+-})\right]\nonumber\\
&+ \left.\ketbra{0}{1}{\eb}\otimes(\cdots) + \ketbra{1}{0}{\eb}\otimes(\cdots)\vphantom{\frac{1}{2}}\right\},
\end{align}
where
\begin{align}
\rho_{\pm\pm} &= \ketbra{\pm t_A\ao}{\pm t_A\ao}{A'}\otimes \ketbra{\pm r_A\ao}{\pm r_A\ao}{\ea},\nonumber\\
\rho_{\pm\mp} &= \ketbra{\pm t_A\ao}{\mp t_A\ao}{A'}\otimes \ketbra{\pm r_A\ao}{\mp r_A\ao}{\ea}.
\end{align}
The states indicated by the triple dots are not important, because when tracing out the auxiliary mode $\eb$ from the state $\rho_{A'\ea B' \eb} $, the terms with $\ketbra{0}{1}{\eb}$ and $\ketbra{1}{0}{\eb}$ are removed.
The tracing of $\eb$ also equates $\ketbra{1}{1}{\eb}$ and $\ketbra{0}{0}{\eb}$ to unity. The tracing of $\ea$ can be done by applying the integrations in Eq.~(\ref{trace_cv}), which transforms $\rho_{\pm\pm}$ and $\rho_{\pm\mp}$ to
\begin{align}
\rho'_{\pm\pm} &= \ketbra{\pm t_A\ao}{\pm t_A\ao}{A'},\nonumber\\
\rho'_{\pm\mp} &= \ketbra{\pm t_A\ao}{\mp t_A\ao}{A'}e^{-2r_A^2 \ao^2},
\end{align}
respectively. After tracing out $\ea$ and $\eb$, we can find the two-mode directly distributed state
\begin{align}
&\rho_{A'B'} = \tr_{\ea\eb} \left[\rho_{A'\ea B'\eb}\right] = \frac{1}{2} \left\{\vphantom{\frac{r_B^2}{N_+^2}}\right.\nonumber\\
&\qquad\; \; \frac{r_B^2}{N_+^2} \; \ketbra{0}{0}{B'}\otimes(\rho'_{++}+\rho'_{--}+\rho'_{-+}+\rho'_{+-})\nonumber\\
&\quad+\;\frac{t_B^2}{N_+^2}\; \ketbra{1}{1}{B'}\otimes(\rho'_{++}+\rho'_{--}+\rho'_{-+}+\rho'_{+-})\nonumber\\
&\quad+\;\frac{1}{N_-^2}  \;   \ketbra{0}{0}{B'}\otimes(\rho'_{++}+\rho'_{--}-\rho'_{-+}-\rho'_{+-})\nonumber\\
&+\frac{t_B}{N_+ N_-} \ketbra{0}{1}{B'}\otimes(\rho'_{++}-\rho'_{--}+\rho'_{-+}-\rho'_{+-})\nonumber\\
&+\frac{t_B}{N_+ N_-} \ketbra{1}{0}{B'}\otimes(\rho'_{++}-\rho'_{--}-\rho'_{-+}+\rho'_{+-})\left.\vphantom{\frac{t_B}{N_+ N_-}}\right\}.
\end{align}
By rearranging the terms, we come to the directly-distributed state given by Eq.~(\ref{arbitrary_alpha}).

\end{appendices}
\end{document}